\documentclass[reprint,amsmath,amssymb,aps,pre,]{revtex4-2}

\usepackage{graphicx}
\usepackage{dcolumn}
\usepackage{bm}
\usepackage{amsthm}

\newtheorem{theorem}{Theorem}
\newtheorem{lemma}[theorem]{Lemma}
\newtheorem{corollary}[theorem]{Corollary}

\theoremstyle{definition}
\newtheorem{definition}[theorem]{Definition}

\begin{document}

\preprint{APS/123-QED}

\title{Statistical analysis of crossing beam speckle fields}

\author{Ian D. Min-Roberts}%

\author{Wojciech Rozmus}%
\affiliation{Department of Physics, University of Alberta, Edmonton, Alberta, Canada}

\author{Pierre A. Michel}%
\affiliation{Lawrence Livermore National Laboratory, Livermore, California, USA}

\date{\today}

\begin{abstract}
Building on the statistical analysis of single beam speckles, this paper develops a statistical description of speckle fields formed by crossing laser beams. A formal definition of individual speckles and their domains allows us to quantify the effective size and abundance of speckles. For two crossing beams, the effective speckle size is derived from the local curvature of the intensity field, and the expected speckle abundance is found asymptotically with the Kac-Rice formula. Monte Carlo simulations of speckle abundance show excellent agreement with the theoretical abundance formula. The two beam speckle analysis is then applied to a symmetric quad beam configuration motivated by polarization smoothing and demonstrates that increasing the separation between beams reduces expected stimulated Brillouin scattering reflectivity from saturated speckles. The statistical model is then extended to an arbitrary number of beams using a coordinate free formulation to obtain the spectral density of laser driven electron density fluctuations.
\end{abstract}


\maketitle


\section{\label{sec:Introduction}Introduction}
A major limitation in inertial confinement fusion comes from laser-plasma instabilities, which can reduce the efficiency with which laser energy is coupled to the target. Optical smoothing techniques are therefore used to mitigate these instabilities by modifying the laser beam in the near field to control the spatial structure at the focal spot. Examples of such techniques include random phase plates (RPPs) \cite{kato1984random}, which break the beam into a fixed speckle pattern, and polarization smoothing \cite{tsubakimoto1992suppression}, which reduces spatial intensity fluctuations by overlapping orthogonally polarized speckle beams. While the techniques reduce instability growth, instabilities such as stimulated Brillouin scattering (SBS) can persist within high intensity speckles. Understanding the statistical properties of these speckles is therefore important for describing laser-plasma instability behaviour in smoothed beams. 

While the statistics of speckles produced by individual beams have been studied extensively \cite{goodman1975statistical,rose1993statistical,garnier1999statistics,michel2023introduction,min2026statistical}, inertial confinement fusion experiments commonly encounter multiple crossing laser beams. In this paper, we therefore consider the statistics of speckles formed by crossing laser beams. When multiple speckle beams overlap, the intensity field depends on the properties of the individual beams as well as their relative polarization and crossing geometry. Polarization determines the extent to which the fields interfere, while the crossing geometry introduces additional spatial scales that influence the size and shape of the speckles \cite{le2011statistical}.

The paper begins by developing a statistical theory for two crossing speckle beams by deriving the intensity probability density function (PDF), and determining the effective speckle sizes and expected abundance as functions of crossing angle and polarization. The expected speckle abundance is obtained asymptotically using the Kac-Rice formula, and compared with Monte Carlo simulations. The two beam result is then applied to a symmetric quad beam configuration based on polarization smoothing at the National Ignition Facility (NIF) and we examine how increasing the angular separation between beams affects the mean SBS reflectivity from individually saturated speckles. We then extend the intensity field analysis to an arbitrary number of crossing beams by deriving a coordinate free expression for the second order statistics of density fluctuations and comparing it with that of a single beam.

\subsection{\label{sec:Speckle_definition}Speckle definition}
A local maximum of the intensity field identifies the centre of a speckle, but does not determine its spatial domain. Geometrically, we can think of a speckle as the intensity hill surrounding a local maximum, consisting of all points whose paths along the intensity gradient lead to that maximum. This allows us to distinguish individual speckles and their spatial domains without imposing any assumption on their shape.  The following definition formalizes this construction.
\begin{definition}[Speckles]\label{def:speckles}
Let $U:\mathbb{R}^N\to[0,\infty)$ be a nonnegative random intensity field with realizations that are at least twice differentiable, and let $\mathbf{x}_*\in\mathbb{R}^N$ be a nondegenerate local maximum for a realization of $U$, so that $\nabla U(\mathbf{x}_*)=0$ and $\nabla\nabla U(\mathbf{x}_*)\prec0$, where $\nabla\nabla$ denotes the Hessian operator and $\prec0$ denotes negative definiteness. We define the speckle associated with $\mathbf{x}_*$ as the set of ordered pairs
\begin{equation}
\mathfrak{S}(\mathbf{x}_*) := \left\{\left(\mathbf{x},U(\mathbf{x})\right): \mathbf{x}\in\Omega_{\mathfrak{S}}(\mathbf{x}_*)\right\},
\label{eq:speckle_definition}
\end{equation}
where $\Omega_{\mathfrak{S}}(\mathbf{x}_*)\subset\mathbb{R}^N$ is the domain of the speckle, taken to be the gradient-ascent basin of $\mathbf{x}_*$, consisting of all points whose gradient-ascent paths converge to $\mathbf{x}_*$.
\end{definition}
The definition is sufficiently general to describe localized intensity structures arising from different physical mechanisms, including speckle patterns in off-axis scattered light \cite{borcea2018laser} and scattered light from biological cells \cite{su20072d}. It also provides a basis for defining characteristic speckle sizes from the local geometry near their maxima and a typical statistical size from an ensemble of speckles. Here, however, we restrict our attention to speckles generated by optical smoothing of laser beams.

\subsection{\label{sec:Speckle_sizes}Speckle sizes}
From Definition~\ref{def:speckles}, a speckle $\mathfrak{S}(\mathbf{x}_*)$ associated with the local maximum $\mathbf{x}_*\in\mathbb{R}^N$ of $U\colon\mathbb{R}^N\to[0,\infty)$ has a domain $\Omega_{\mathfrak{S}}(\mathbf{x}_*)$, from which measures of its size could in principle be obtained and averaged over many maxima. Here, however, we are interested in its local behaviour and thus the local size of the speckle near its peak. We therefore introduce an effective speckle size determined analytically from the curvature of the intensity field about $\mathbf{x}_*$. For a small displacement $\mathbf{x}^\prime = \mathbf{x}-\mathbf{x}_*$ from $\mathbf{x}_*$, the Taylor expansion of the intensity field around the maximum $\mathbf{x}_*$ can be expressed as
\[
U(\mathbf{x}) = U(\mathbf{x}_*) + \frac{1}{2}{\mathbf{x}^{\prime}}^{\top}\nabla\nabla U(\mathbf{x}_*)\mathbf{x}^\prime + o\left(\lvert \mathbf{x}^\prime\rvert^2\right),
\]
where $o(\cdot)$ denotes terms that are asymptotically smaller than their argument, and the superscript $\top$ denotes transposition. Dividing by the peak intensity $U(\mathbf{x}_*)$ gives
\begin{equation}
\frac{U(\mathbf{x})}{U(\mathbf{x}_*)} = 1 + \frac{{\mathbf{x}^{\prime}}^{\top}\nabla\nabla U(\mathbf{x}_*)\mathbf{x}^\prime}{2U(\mathbf{x}_*)} + o\left(\lvert \mathbf{x}^\prime\rvert^2\right).
\label{eq:speckle_size}
\end{equation}
This expansion shows that the effective local size of an individual speckle can be characterized by its normalized curvature $-\nabla\nabla U(\mathbf{x}_*)/2U(\mathbf{x}_*)$. While this quantity varies from one speckle to another, its high intensity asymptotic behaviour can be characterized by an effective second order spectral moment $\boldsymbol{\Lambda}$ such that
\begin{equation}
-\frac{\nabla\nabla U(\mathbf{x}_*)}{2U(\mathbf{x}_*)} = \boldsymbol{\Lambda} + O_{\mathrm{p}}\left(U(\mathbf{x}_*)^{-1/2}\right),
\label{eq:effective_second_moment}
\end{equation}
where $O_{\mathrm{p}}(\cdot)$ denotes stochastic terms of the stated order in probability. For the nondegenerate beam spectra considered here, $\boldsymbol{\Lambda}\succ0$ (is positive definite). Furthermore, since the intensity is nonnegative by definition, we take the zero level of the local quadratic approximation in Eq.~\eqref{eq:speckle_size} as a measure of the effective spatial extent of a speckle. Thus, combining Eqs.~\eqref{eq:speckle_size} and~\eqref{eq:effective_second_moment}, the effective size of a speckle is described by
\begin{equation}
{\mathbf{x}^{\prime}}^{\top} \boldsymbol{\Lambda} \mathbf{x}^{\prime} = 1,
\label{eq:speckle_ellipsoid}
\end{equation}
which defines an ellipsoid with characteristic lengths $\ell_j$ determined by the eigenvalues $\lambda_j$ of $\boldsymbol{\Lambda}$ such that
\begin{equation}
\ell_j=\lambda_j^{-1/2}, \qquad \text{for all } j\in[1,N].
\label{eq:speckle_lengths}
\end{equation}

The effective second order spectral moment can be calculated explicitly for a single speckle beam, allowing us to recover the characteristic lengths obtained in Ref.~\cite{min2026statistical}. For this calculation, we take the normalized intensity in terms of the complex vector potential envelope $\mathcal{A}\colon\mathbb{R}^N\to\mathbb{C}$ as
\[
U(\mathbf{x}) = \frac{1}{\sigma^2}\lvert\mathcal{A}(\mathbf{x})\rvert^2, \qquad \mathcal{A}\sim\mathcal{CN}(0,\sigma^2).
\]
The Hessian of the normalized intensity is
\[
\nabla\nabla U = \frac{2}{\sigma^2}\, \operatorname{Re}\!\left\{\left(\nabla\nabla \!\mathcal{A}\right)\,\overline{\mathcal{A}} + \nabla\!\mathcal{A}\,\nabla\!\mathcal{A}^{\dagger}\right\},
\]
where the overline denotes complex conjugate and the superscript $\dagger$ denotes the conjugate transpose. To simplify the curvature calculation, we remove the linear phase associated with the first order spectral moment by defining the phase corrected envelope
\begin{equation}
\mathring{\mathcal{A}}(\mathbf{x}) = \mathcal{A}(\mathbf{x})\, e^{-i\mathbf{m}^{(1)}\cdot\mathbf{x}},
\label{eq:phase_corrected_envelope}
\end{equation}
where $\mathbf{m}^{(j)}$ denotes the normalized $j$th spectral moment of $\mathcal{A}$,
\begin{equation}
\mathbf{m}^{(j)} = \frac{1}{\sigma^2}\int_{\mathbb{R}^N}\mathbf{k}^{\otimes j} S_{\mathcal{A}\mathcal{A}}(\mathbf{k})\, d\mathbf{k}.
\label{eq:jth_order_spectral_moments}
\end{equation}
Here, $S_{\mathcal{A}\mathcal{A}}$ is the spectral density of $\mathcal{A}$, and $\mathbf{k}^{\otimes j}$ denoting the $j$-fold tensor product of the spatial wavevector $\mathbf{k}$. This transformation leaves $\lvert\mathcal{A}\rvert^2$, and hence $U$ and its Hessian, unchanged~\cite{min2026statistical}. Gaussian conditioning then gives $\langle \nabla\nabla \mathring{\mathcal{A}} \mid \mathring{\mathcal{A}} \rangle = -\mathring{\mathbf{m}}^{(2)}\mathring{\mathcal{A}}$, where
\begin{equation}
\mathring{\mathbf{m}}^{(2)}:=\mathbf{m}^{(2)}-\mathbf{m}^{(1)}\otimes\mathbf{m}^{(1)}.
\label{eq:phase_shifted_second_order_spectral_moment}
\end{equation}
Fluctuations of $\nabla\nabla \mathring{\mathcal{A}}$ about its conditional mean remain $O_{\mathrm{p}}(1)$, while the gradient term contributes only at lower order. Therefore, at high intensity maxima,
\[
-\frac{\nabla\nabla U(\mathbf{x}_*)}{2U(\mathbf{x}_*)} = \mathring{\mathbf{m}}^{(2)} + O_{\mathrm{p}}\left(U(\mathbf{x}_*)^{-1/2}\right).
\]
Comparison with Eq.~\eqref{eq:effective_second_moment} therefore gives $\boldsymbol{\Lambda} = \mathring{\mathbf{m}}^{(2)}$, which is consistent with the single beam result from Ref.~\cite{min2026statistical}. The dimension $N$ is determined by the domain of the vector potential envelope $\mathcal{A}$. For RPP speckles, $N=3$ describes the full spatial field, while $N=2$ is the restriction to a plane. For a plane spanned by two orthonormal vectors $\hat{\mathbf{e}}_1$ and $\hat{\mathbf{e}}_2$, the corresponding effective second order spectral moment is given by
\begin{equation}
\begin{pmatrix}
\hat{\mathbf{e}}_1 & \hat{\mathbf{e}}_2
\end{pmatrix}^\top \boldsymbol{\Lambda}
\begin{pmatrix}
\hat{\mathbf{e}}_1 & \hat{\mathbf{e}}_2
\end{pmatrix}
\label{eq:Lambda_projection}
\end{equation}
where $\boldsymbol{\Lambda}$ is the three-dimensional moment. For temporal smoothing techniques, such as induced spatial incoherence (ISI)~\cite{lehmberg1987theory}, the full spatiotemporal field has $N=4$, with lower dimensional restrictions obtained by a projection analogous to Eq.~\eqref{eq:Lambda_projection}.

\subsection{\label{sec:Speckle_abundance}Speckle abundance}
From Definition~\ref{def:speckles}, each speckle is associated with a unique local intensity maximum. Thus, the number of speckles with peak intensity exceeding $u$ is simply the number of local maxima above $u$. For a stationary random intensity field $U$ in a compact domain $\Omega\subset\mathbb{R}^N$, the Kac-Rice theorem~\cite[Theorem 11.2.1]{adler2007random} gives the expected number of such maxima, provided its regularity and nondegeneracy conditions are satisfied,
\begin{equation}
\begin{aligned}
\left\langle M_{\Omega}^{+}(u)\right\rangle ={}& \lvert\Omega\rvert\,\Big\langle\lvert\det(\nabla\nabla U)\rvert\,\mathbf{1}_{\{U>u\}}\\
&\times\mathbf{1}_{\left\{\nabla\nabla U\prec0\right\}}\,\Big|\,\nabla U=0\Big\rangle f_{\nabla U}(0),
\end{aligned}
\label{eq:kac_rice}
\end{equation}
where $\mathbf{1}_{X}$ denotes the indicator function on a set $X$, and $f_{\nabla U}$ denotes the probability density function of $\nabla U$. The single beam speckle abundance formula derived in Ref.~\cite{min2026statistical} follows from a theorem based on Eq.~\eqref{eq:kac_rice}. Earlier single beam abundance formulas derived by Rose and DuBois~\cite{rose1993statistical} and later corrected by Garnier~\cite{garnier1999statistics} also trace back to this equation, through an intermediate result that required an additional ansatz.

\section{\label{sec:Two_Crossing_Beams}Two crossing beams}
\begin{figure}[b]
\includegraphics[width=\linewidth]{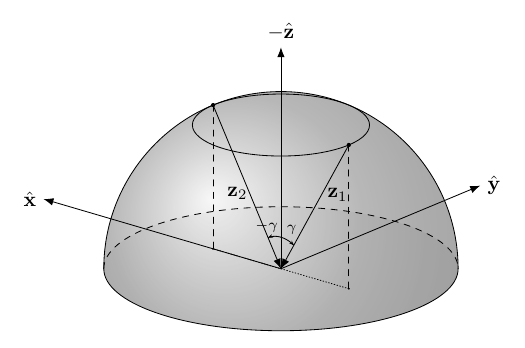}
\caption{The geometry of the two crossing beam configuration. The propagation directions $\mathbf{z}_1$ and $\mathbf{z}_2$ make angles $\gamma$ with the global $\hat{\mathbf{z}}$ axis.}
\label{fig:double_crossing_beams} 
\end{figure}

Crossing speckle beams introduces a unique spatial structure arising from their relative polarization and propagation directions that are absent in single beams~\cite{le2011statistical}. The two beam configuration allows these effects to be examined transparently. So consider two identical RPP speckle beams crossing symmetrically about the global $\hat{\mathbf{z}}$ axis in the $xz$-plane, each with its own linear polarization direction, as illustrated in Fig.~\ref{fig:double_crossing_beams}. Each beam propagates at an angle $\gamma$ with respect to $\hat{\mathbf{z}}$, while the aperture may be either circular or square.

\subsection{\label{sec:Vector_potential_2}Vector potential}
The total complex vector potential $\mathbf{A}\colon\mathbb{R}^3\times\mathbb{R}_+\to\mathbb{C}^3$ of the two intersecting beams is the superposition of the individual beam fields $\mathbf{A}_j\colon\mathbb{R}^3\times\mathbb{R}_+\to\mathbb{C}^3$
\begin{equation}
\mathbf{A}(x,y,z,t) = \mathbf{A}_1(x,y,z,t) + \mathbf{A}_2(x,y,z,t),
\label{eq:two_beam_field}
\end{equation}
where $(x,y,z)\in\mathbb{R}^3$ denotes the spatial coordinate system spanning the set of global unit vectors $\{\hat{\mathbf{x}},\hat{\mathbf{y}},\hat{\mathbf{z}}\}$, and $t\in\mathbb{R}_+$ denotes time. We define a local coordinate system associated with beam $j$ as $(x_j,y_j,z_j)$ spanning the orthonormal basis $\{\hat{\mathbf{x}}_j,\hat{\mathbf{y}}_j,\hat{\mathbf{z}}_j\}$, such that $\hat{\mathbf{z}}_j$ is aligned with the propagation direction of the beam and $\hat{\mathbf{y}}_j$ is aligned with the polarization direction. Under the paraxial approximation, the complex vector potential of beam $j$ may be written as
\begin{equation}
\mathbf{A}_j(x,y,z,t)=\mathring{\mathcal{A}}_j(x_j,y_j,z_j)e^{i\psi(z_j,t)}\hat{\mathbf{y}}_j,
\label{eq:paraxial_beam}
\end{equation}
where $\mathring{\mathcal{A}}_j\colon\mathbb{R}^3\to\mathbb{C}$ is the phase corrected envelope defined in Eq.~\eqref{eq:phase_corrected_envelope}, and
\begin{equation}
\psi(z_j,t)=\kappa_0z_j-\omega_0t, \qquad \kappa_0:=k_0+m_z^{(1)},
\label{eq:oscillating_phase_term}
\end{equation}
is the oscillating phase term, with $\omega_0$ the carrier frequency, $k_0$ the carrier wavenumber, and $m_z^{(1)}$ the longitudinal component of the first order spectral moment, $\mathbf{m}^{(1)} = m_z^{(1)}\hat{\mathbf{z}}_j$. For example, a beam with a circular aperture has longitudinal first order spectral moment $m_z^{(1)} = -k_c^2/(4k_0)$ and a beam with a square aperture would have $m_z^{(1)} = -k_c^2/(3k_0)$, where
\begin{equation}
k_c = \frac{k_0\tilde{D}}{\sqrt{\tilde{D}^{\,2}+4\tilde{f}^{2}}} \approx \frac{k_0\tilde{D}}{2\tilde{f}}, \qquad \tilde f \gg \tilde D,
\label{eq:wave_number_cutoff}
\end{equation}
for focal length $\tilde{f}$ and aperture diameter $\tilde{D}$~\cite{min2026statistical}. Thus, the total vector potential may be written as
\begin{equation}
\mathbf{A}(x,y,z,t) = \mathring{\mathcal{A}}_1 e^{i\psi(z_1,t)}\,\hat{\mathbf{y}}_1 + \mathring{\mathcal{A}}_2 e^{i\psi(z_2,t)}\,\hat{\mathbf{y}}_2.
\label{eq:two_beam_field_paraxial}
\end{equation}

In general, the local coordinates of beam $j$ are related to the global coordinates through the rotational matrix $\mathbf{R}_{j}\in\mathrm{SO}(3)$ according to
\begin{equation}
\begin{pmatrix}x\\y\\z\end{pmatrix} = \mathbf{R}_{j} \begin{pmatrix}x_j\\y_j\\z_j\end{pmatrix},
\qquad
\begin{pmatrix}x_j\\y_j\\z_j\end{pmatrix} = \mathbf{R}_{j}^\top \begin{pmatrix}x\\y\\z\end{pmatrix}.
\label{eq:coordinate_rotation}
\end{equation}
For the geometry illustrated in Fig.~\ref{fig:double_crossing_beams}, the propagation directions are determined by rotations about the global $\hat{\mathbf{y}}$ axis, while the polarization directions are set by rotations about the respective beam axes. We choose $\hat{\mathbf{y}} = \hat{\mathbf{y}}_1$, while $\hat{\mathbf{y}}_2$ makes an angle $\arccos\alpha$ with respect to $\hat{\mathbf{y}}$ so that $\lvert\hat{\mathbf{y}}_1\cdot\hat{\mathbf{y}}_2\rvert = \alpha$. Thus,
\begin{equation}
\mathbf{R}_{j} = \mathbf{R}_y\left(\pm\gamma\right)\mathbf{R}_z\left((j-1)\arccos\alpha\right),
\label{eq:coordinate_rotation_product}
\end{equation}
where the upper and lower signs correspond to $j=1$ and $j=2$, respectively. In Ref.~\cite{min2026statistical}, the phase corrected second order spectral moment was shown to be isotropic in the transverse plane
\begin{equation}
\mathring{\mathbf{m}}^{(2)} =
\begin{pmatrix}
\lambda_{\perp} & 0 & 0\\
0 & \lambda_{\perp} & 0\\
0 & 0 & \lambda_{\parallel}
\end{pmatrix},
\label{eq:1_spectral_moment}
\end{equation}
thus rotations about the local propagation axis leave $\mathring{\mathbf{m}}^{(2)}$ invariant, which implies the second order spectral moment of beam $j$ in the global coordinate system can be written as
\begin{equation}
\mathring{\mathbf{m}}_j^{(2)} = \mathbf{R}_y(\pm\gamma)\,\mathring{\mathbf{m}}^{(2)}\, \mathbf{R}_y(\pm\gamma)^\top,
\label{eq:spectral_moment_rotation}
\end{equation}
where
\begin{equation}
\mathbf{R}_y\left(\pm\gamma\right) =
\begin{pmatrix}
\cos\gamma & 0 & \pm\sin\gamma\\
0 & 1 & 0\\
\mp\sin\gamma & 0 & \cos\gamma
\end{pmatrix}.
\label{eq:rotation}
\end{equation}
For circular and square flat-top apertures, the eigenvalues $\lambda_\perp$ and $\lambda_\parallel$ of the phase corrected second order spectral moment are given by
\begin{align}
\text{circular:}\qquad & \lambda_\perp \approx \frac{k_0^2}{16f_\#^2}, \qquad \lambda_\parallel \approx \frac{k_0^2}{768f_\#^4}\label{eq:lambda_eigenvalues_circular},\\
\text{square:}\qquad & \lambda_\perp \approx \frac{k_0^2}{12f_\#^2}, \qquad \lambda_\parallel \approx \frac{k_0^2}{360f_\#^4}\label{eq:lambda_eigenvalues_square},
\end{align}
where $f_\#=\tilde{f}/\tilde{D}$ is the f-number of the beam, and the approximation given in Eq.~\eqref{eq:wave_number_cutoff} has been used.

The one-point statistics of the scalar envelope $\mathring{\mathcal A}_j$ can be obtained from the central limit theorem, since each beam contains many independently phased spectral modes \cite{rose1993statistical,garnier1999statistics,min2026statistical}. Thus, at any fixed spatial point,
\[
\mathring{\mathcal{A}}_j \sim \mathcal{CN}(0,\sigma^2), \qquad \sigma^2 := \left\langle\mathring{\mathcal{A}}_j\overline{\mathring{\mathcal{A}}_j}\right\rangle.
\]
Since the total field in Eq.~\eqref{eq:two_beam_field} is a linear combination of independent proper complex Gaussian variables, it is itself a mean-zero proper complex Gaussian vector field,
\begin{equation}
\mathbf{A} \sim \mathcal{CN} \left(\mathbf{0}, \boldsymbol{\Sigma}\right), \quad \boldsymbol{\Sigma} := \left\langle\mathbf{A}\mathbf{A}^{\dagger}\right\rangle,
\label{eq:vector_gaussian_distribution}
\end{equation}
where $\boldsymbol{\Sigma}$ denotes its one-point covariance matrix which, by direct calculation, can be expressed as
\begin{equation}
\boldsymbol{\Sigma} = \sigma^2 \left(\hat{\mathbf{y}}_{1}\otimes\hat{\mathbf{y}}_1+\hat{\mathbf{y}}_{2}\otimes\hat{\mathbf{y}}_2\right).
\label{eq:covariance_polarization_sum}
\end{equation}

\subsection{\label{sec:Intensity_2}Intensity}
\begin{figure}
\includegraphics[width=\linewidth]{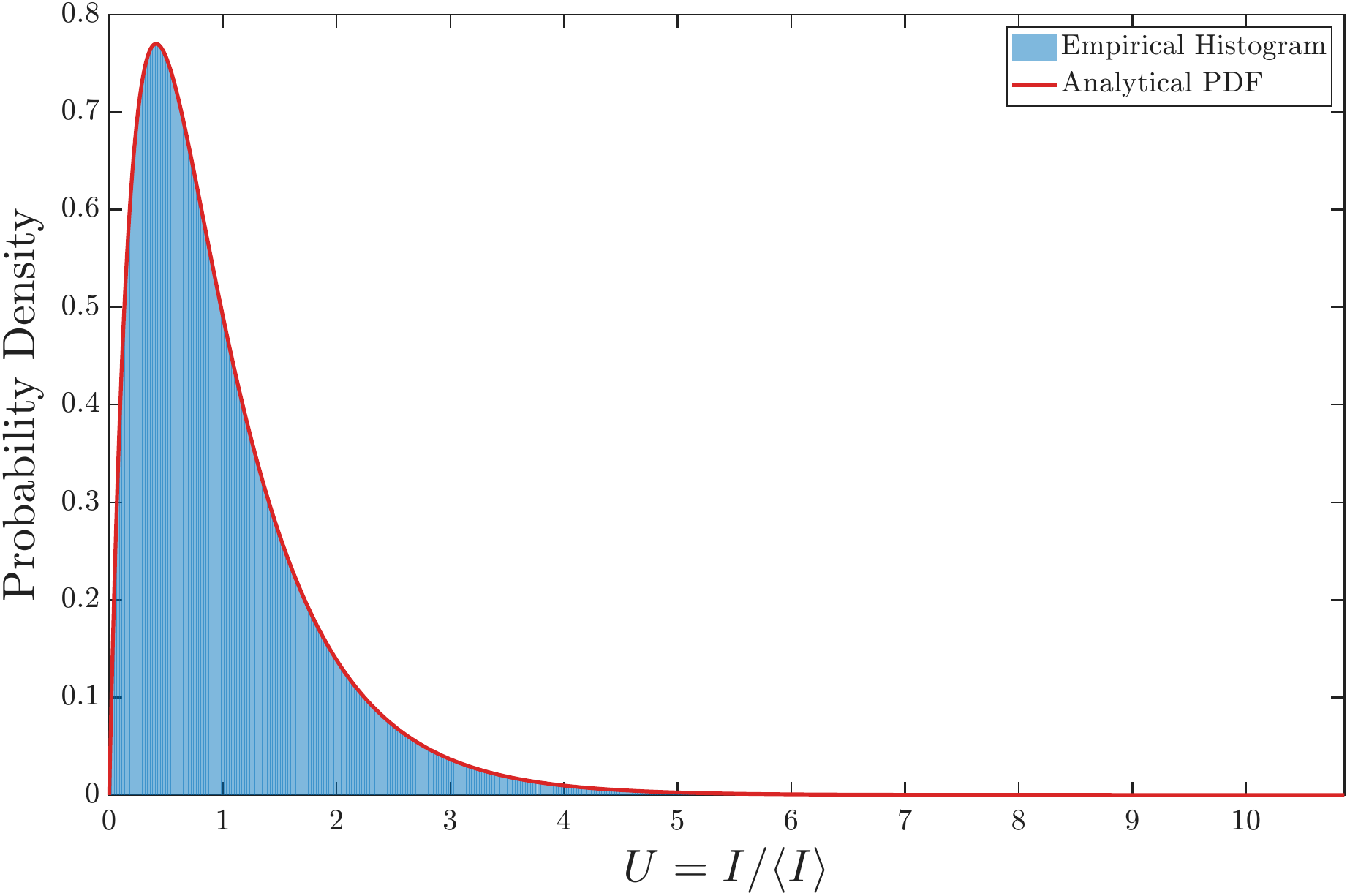}
\caption{Intensity probability density for a single realization of two crossing beams with $\alpha = 1/2$ and $\gamma = 20^\circ$ compared with the theoretical distribution of Eq.~\eqref{eq:pdf_2beam}.}
\label{fig:2_pdf_intensity}
\end{figure}
The intensity field is defined as
\begin{equation}
I := \frac{1}{2}\varepsilon_0c\,\omega_0^2|\mathbf{A}|^2.
\label{eq:intensity_definition}
\end{equation}
The mean intensity of an individual beam and that of the two crossing beam field are respectively
\begin{equation}
I_0 := \frac{1}{2}\varepsilon_0c\,\omega_0^2\sigma^2, \qquad \langle I\rangle=2I_0.
\label{eq:mean_intensity}
\end{equation}
Then from Eq.~\eqref{eq:two_beam_field_paraxial}, the normalized intensity $U:=I/\langle I\rangle$ can be expressed as
\begin{equation}
U = \frac{1}{2\sigma^2} \left[\lvert\mathring{\mathcal{A}}_1\rvert^2 + \lvert\mathring{\mathcal{A}}_2\rvert^2 + 2\alpha\,\operatorname{Re}\!\left\{\overline{\mathring{\mathcal{A}}_1}\mathring{\mathcal{A}}_2 e^{i\Delta\psi}\right\}\right],
\label{eq:crossing_beam_intensity}
\end{equation}
where $\alpha = \lvert\hat{\mathbf{y}}_1\cdot\hat{\mathbf{y}}_2\rvert$ measures the polarization overlap, and $\Delta\psi := \psi(z_2,t) - \psi(z_1,t) = -2\kappa_0x\sin\gamma$ is the crossing beam interference phase. To find its probability distribution, we define
\[
\mathring{\mathcal{A}}_\pm := \frac{1}{\sqrt{2}}\left(\mathring{\mathcal{A}}_1\pm\mathring{\mathcal{A}}_2e^{i\Delta\psi}\right)
\]
which is a unitary transformation at any fixed spatial point, implying
\[
\mathring{\mathcal{A}}_+,\mathring{\mathcal{A}}_-\overset{\mathrm{ind}}{\sim}\mathcal{CN}(0,\sigma^2).
\]
Eq.~\eqref{eq:crossing_beam_intensity} can then be written as
\[
U = \frac{1+\alpha}{2\sigma^2} \lvert\mathring{\mathcal{A}}_+\rvert^2 + \frac{1-\alpha}{2\sigma^2} \lvert\mathring{\mathcal{A}}_-\rvert^2,
\]
and since $2\lvert\mathring{\mathcal{A}}_\pm\rvert^2\sim\sigma^2\chi_2^2$, this is a rank 2 weighted sum of independent $\chi_2^2$ random variables with coefficients $w_{\pm}/2$, where
\[
w_{\pm} = \frac{1\pm\alpha}{2}.
\]
The general rank $r$ result for such a distribution is derived in Appendix~\ref{sec:Probability_density_function}. Applying that result here gives
\begin{equation}
f_U(u)=\frac{1}{\alpha}\left[e^{-2u/(1+\alpha)}-e^{-2u/(1-\alpha)}\right],
\label{eq:pdf_2beam}
\end{equation}
for $u\geq0$ and $0<\alpha<1$ with the limiting cases $\alpha\to0$ and $\alpha\to1$ giving $f_{U}(u)=4ue^{-2u}$ and $f_{U}(u)=e^{-u}$, respectively. Fig.~\ref{fig:2_pdf_intensity} shows that, unlike for a single beam~\cite{min2026statistical}, the most commonly occurring intensity is not necessarily zero.

\subsection{\label{sec:Speckles_2}Speckles}
We turn our attention to the statistics of speckles. We first determine the effective size of a speckle, then give an asymptotic formula for their abundance above an intensity level set.

\subsubsection{Speckle sizes}
\begin{figure}
\includegraphics[width=\linewidth]{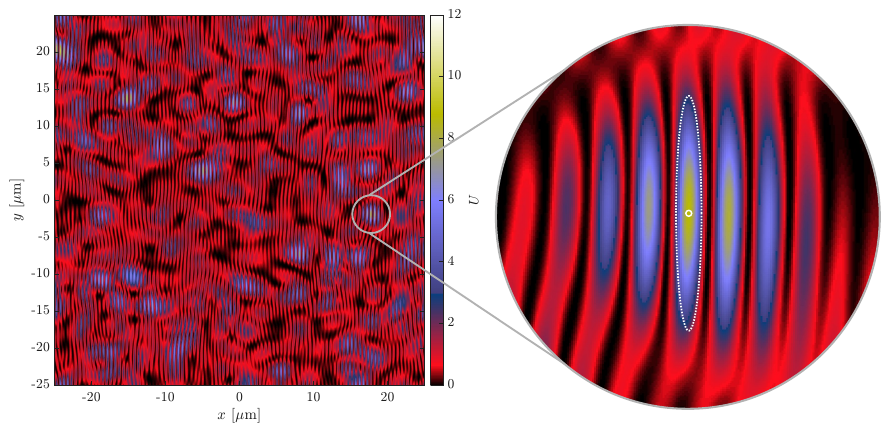}
\caption{A simulation of the normalized intensity field for two crossing speckle beams described by Eq.~\eqref{eq:crossing_beam_intensity} projected in the $(x,y)$ plane. The polarization overlap was set to $\alpha=0.9$, and crossing angle was set to $\gamma=30^\circ$. The circular inset shows an enlarged view of a cluster of speckles. The dashed ellipse is the $(x,y)$ cross-section of the local quadratic zero level surface described in Eq.~\eqref{eq:two_beam_speckle_ellipsoid}, with $\ell_x$ and $\ell_y$ given by Eq.~\eqref{eq:two_beam_speckle_lengths}.}
\label{fig:SpeckleSizeSimFig}
\end{figure}

Applying the formulas used to obtain effective speckle sizes from Section~\ref{sec:Speckle_sizes} to our two crossing beam model gives the following lemma.
\begin{lemma}[Effective Second Order Spectral Moment for Two Crossing Beams]\label{lem:speckle_size}
Let $U:\mathbb{R}^3\to[0,\infty)$ be described by Eq.~\eqref{eq:crossing_beam_intensity}, and let $\mathbf{x}_*$ be a high-intensity local maximum of $U$. For polarization overlap $0<\alpha\leq1$, the effective second-order spectral moment introduced in Eq.~\eqref{eq:effective_second_moment} is
\begin{equation}
\boldsymbol{\Lambda} = \frac{1}{2}\left[\mathring{\mathbf{m}}_1^{(2)}+\mathring{\mathbf{m}}_2^{(2)}+\frac{4\alpha\kappa_0^2}{1+\alpha}\sin^2\gamma\,\hat{\mathbf{x}}\otimes\hat{\mathbf{x}}\right].
\label{eq:2_effective_second_order_spectral_moment}
\end{equation}
For $\alpha=0$, we parameterize the relative contributions of the two beams at $\mathbf{x}_*$ by $|\mathring{\mathcal{A}}_1(\mathbf{x}_*)|^2=2\sigma^2U(\mathbf{x}_*)\sin^2\vartheta$ and $|\mathring{\mathcal{A}}_2(\mathbf{x}_*)|^2=2\sigma^2U(\mathbf{x}_*)\cos^2\vartheta$. The effective second-order spectral moment in Eq.~\eqref{eq:effective_second_moment} is then
\begin{equation}
\boldsymbol{\Lambda}(\vartheta):=\sin^2\vartheta\,\mathring{\mathbf{m}}_1^{(2)}+\cos^2\vartheta\,\mathring{\mathbf{m}}_2^{(2)}.
\label{eq:two_beam_effective_moment_alpha0}
\end{equation}
So, for $\alpha=0$, the effective second-order spectral moment depends on $\vartheta$ and therefore on the relative contributions of the two beams at the maximum.
\begin{proof}
See Appendix~\ref{sec:Proof_of_Lemma}.
\end{proof}
\end{lemma}
For $0<\alpha\leq1$, the speckle lengths can be determined along principal directions coinciding with the global coordinates using Eq.~\eqref{eq:speckle_lengths}, namely
\begin{equation}
\begin{aligned}
\ell_x &= \left(\lambda_\perp\cos^2\gamma+\lambda_\parallel\sin^2\gamma+\frac{2\alpha\kappa_0^2}{1+\alpha}\sin^2\gamma\right)^{-1/2},\\
\ell_y &= \lambda_\perp^{-1/2}, \quad \ell_z = \left(\lambda_\perp\sin^2\gamma+\lambda_\parallel\cos^2\gamma\right)^{-1/2},
\end{aligned}
\label{eq:two_beam_speckle_lengths}
\end{equation}
where $\lambda_\perp$ and $\lambda_\parallel$ are defined from the phase corrected second order spectral moment of a single beam in Eq.~\eqref{eq:1_spectral_moment}, with explicit expressions for circular and square flat-top apertures given in Eqs.~\eqref{eq:lambda_eigenvalues_circular} and \eqref{eq:lambda_eigenvalues_square}, respectively. The zero level set of the quadratic approximation defined in Eq.~\eqref{eq:speckle_ellipsoid} therefore takes the form
\begin{equation}
\frac{(x-x_*)^2}{\ell_x^2}+\frac{(y-y_*)^2}{\ell_y^2}+\frac{(z-z_*)^2}{\ell_z^2}=1,
\label{eq:two_beam_speckle_ellipsoid}
\end{equation}
which is an ellipsoid centered at the speckle maximum $(x_*,y_*,z_*)$ with principal semi-axis $\ell_x$, $\ell_y$, and $\ell_z$. The crossing beam geometry alters speckle lengths in the $x$ and $z$ direction though rotated single beam moments, while the interference from the beams contributes only in the $x$ direction; the speckle length in the $y$ direction remains unchanged. A cross-section of the ellipsoid in the $(x,y)$ plane is shown in Fig.~\ref{fig:SpeckleSizeSimFig}

\begin{figure*}
\centering
\includegraphics[width=0.33\textwidth]{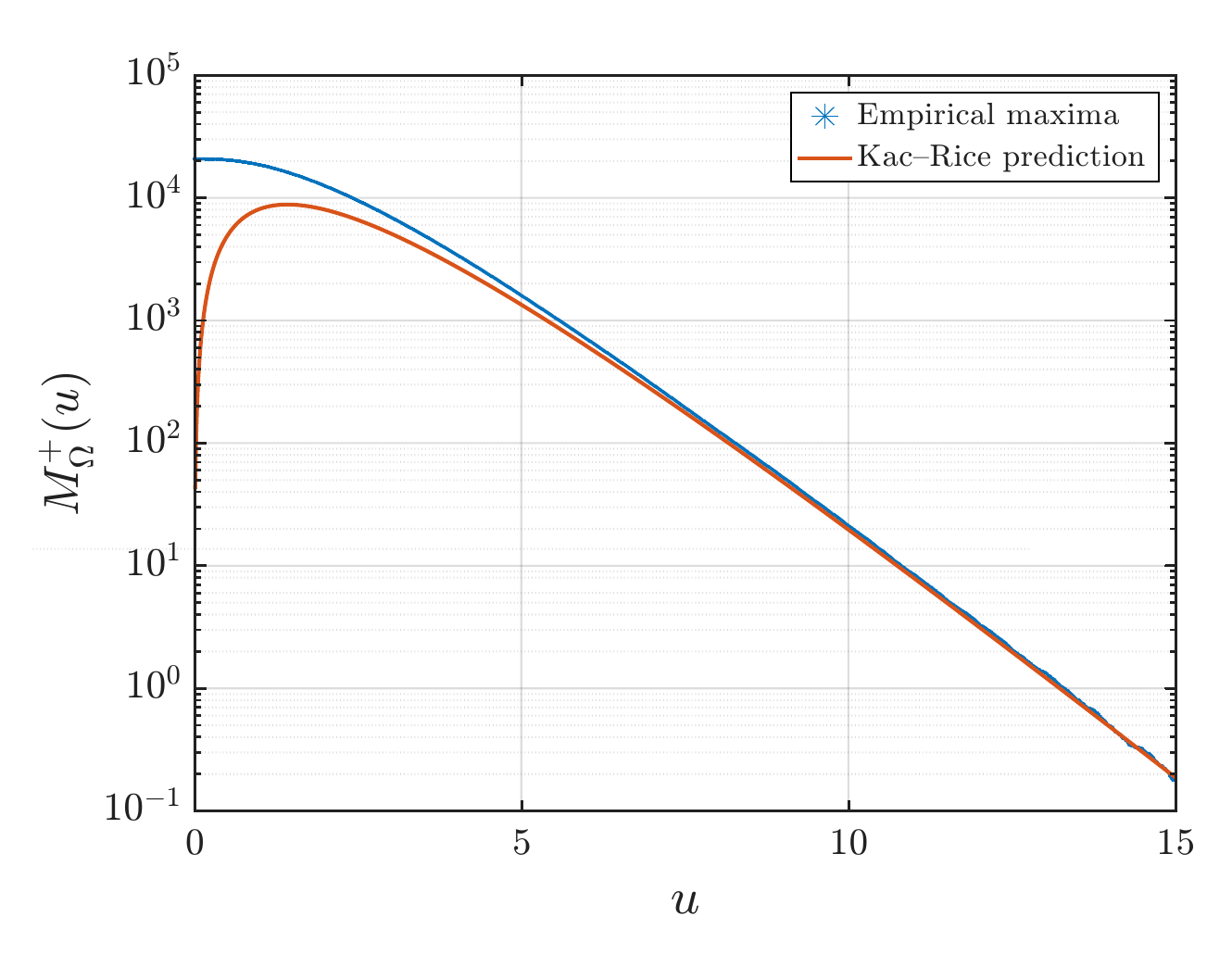}\hfill
\includegraphics[width=0.33\textwidth]{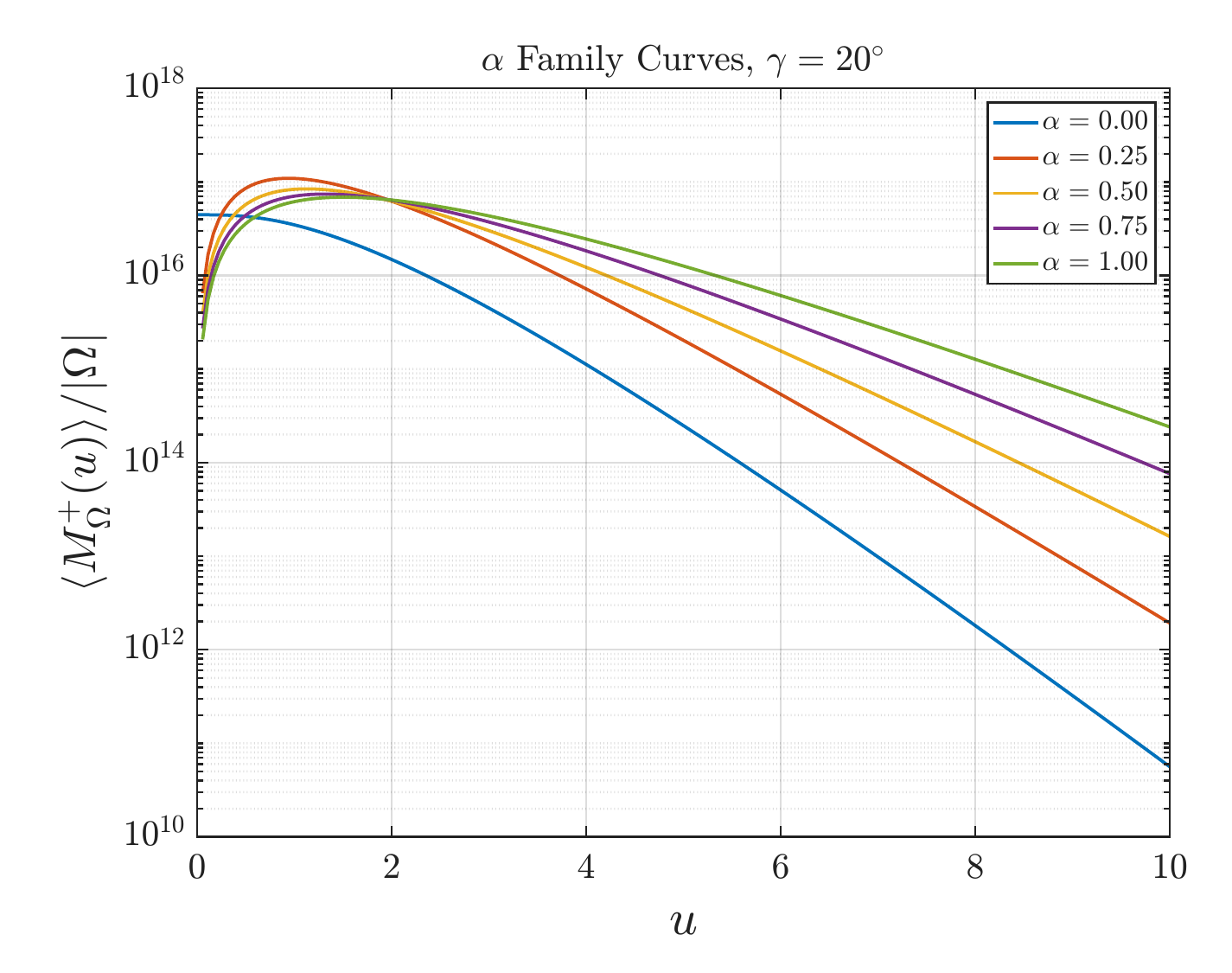}\hfill
\includegraphics[width=0.33\textwidth]{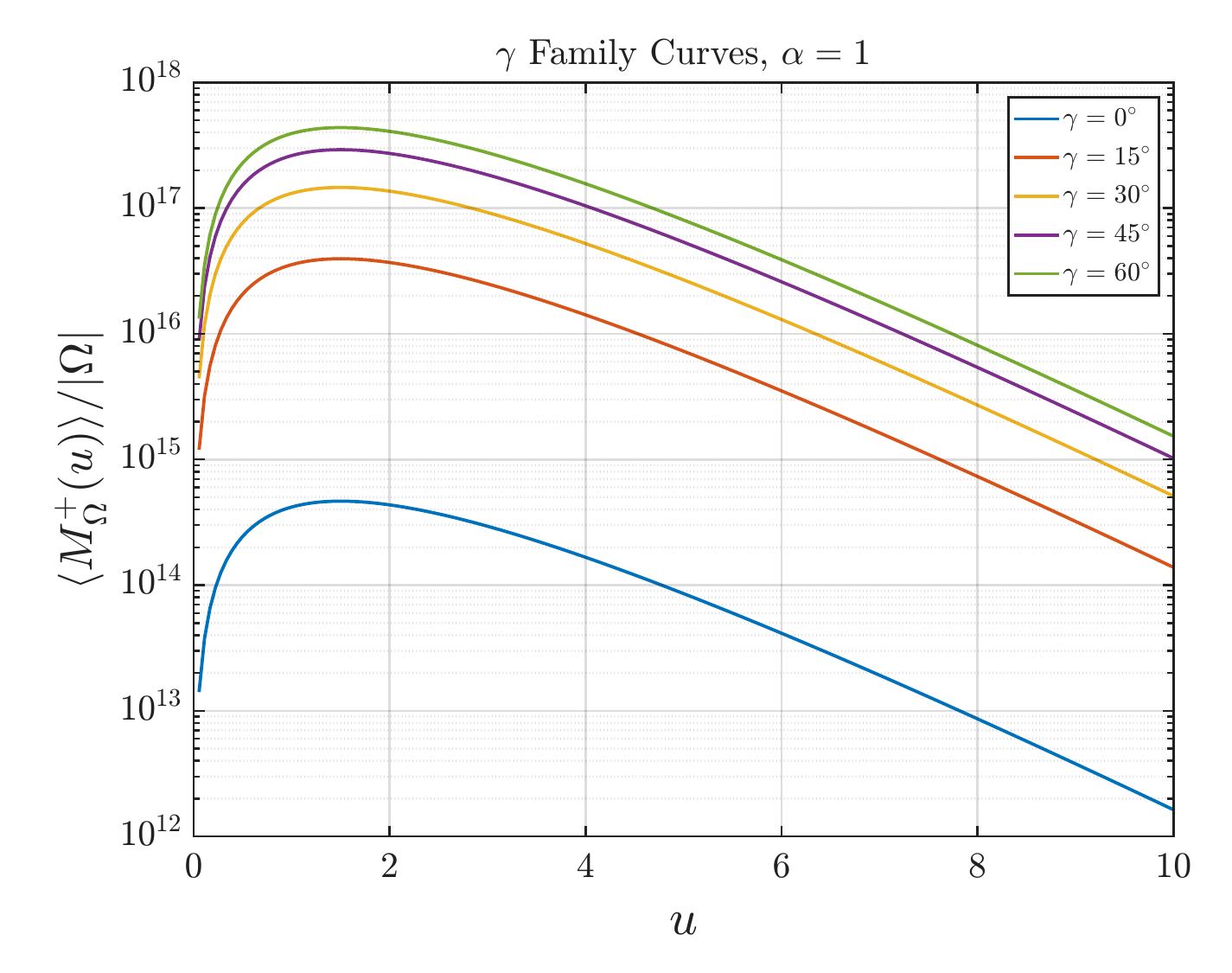}
\caption{Expected number of speckles exceeding an intensity level set $u$ for two crossing beams, given by Theorem~\ref{the:speckle_abundance}. Left: Comparison of Eq.~\eqref{eq:maxima_0alpha1} with Monte Carlo simulations using 200 realizations of two crossing RPP beams, each with a circular flat-top aperture containing 208 RPP phase elements, $f$-number $6.7$, and wavelength $351\,\mathrm{nm}$. The polarization overlap is $\alpha=0.9$, and the beam angle is $\gamma=20^\circ$. Middle: Family of curves showing the expected speckle abundance for varying polarization overlap $\alpha$ at fixed beam angle $\gamma=20^\circ$. Right: Corresponding family of curves for varying beam angle $\gamma$ at fixed polarization overlap $\alpha=1$.}
\label{fig:speckle_abundance_results}
\end{figure*}

\subsubsection{Speckle abundance}
Ref.~\cite{min2026statistical} obtained the expected number of local maxima above a given level set for a single beam by applying results from Worsley~\cite{worsley1994local}. However, for the two beam configuration, the intensity field no longer satisfies the assumptions underlying Worsley's results, and no directly applicable formula is available~\footnote{The Gaussian kinematic formula from Taylor~\cite{taylor2006gaussian} is also not directly applicable, as it assumes independent, identically distributed Gaussian component fields, whereas the crossing beams generally have different spatial covariance functions.}. We therefore start from the general Kac-Rice expression given in Eq.~\eqref{eq:kac_rice} and derive the corresponding high threshold asymptotics.
\begin{theorem}[Number of Speckles Above a Given Level Set for Two Crossing Beams]\label{the:speckle_abundance}
For $N=2,3$, let $U:\mathbb{R}^N\to[0,\infty)$ be the normalized intensity field defined in Eq.~\eqref{eq:crossing_beam_intensity}, restricted to a plane when $N=2$, and let $M_\Omega^+(u)$ denote the number of speckles, as defined in Definition~\ref{def:speckles}, with maxima $\mathbf{x}_*\in\Omega$ satisfying $U(\mathbf{x}_*)>u$. For two identical crossing beams with polarization overlap $0<\alpha\leq1$, as $u\to\infty$,
\begin{equation}
\begin{aligned}
\left\langle M_\Omega^+(u)\right\rangle ={}& \frac{|\Omega|\det(\boldsymbol{\Lambda})^{\frac{1}{2}}}{\pi^{N/2}}\frac{1+\alpha}{2\alpha}\left(\frac{2u}{1+\alpha}\right)^{N/2}\\
&\times e^{-2u/(1+\alpha)}\left[1+O(u^{-1/2})\right],
\end{aligned}
\label{eq:maxima_0alpha1}
\end{equation}
where $\boldsymbol{\Lambda}$ is the effective second order spectral moment defined in Lemma~\ref{lem:speckle_size}, Eq.~\eqref{eq:2_effective_second_order_spectral_moment} and projected onto the corresponding plane using Eq.~\eqref{eq:Lambda_projection} when $N=2$. For $\alpha=0$,
\begin{equation}
\begin{aligned}
\left\langle M_\Omega^+(u)\right\rangle ={}& \frac{2^{\frac{N}{2}+1}|\Omega|}{\pi^{N/2}} \int_0^{\frac{\pi}{2}}\sin(2\vartheta)\sqrt{\det\boldsymbol{\Lambda}(\vartheta)}\,d\vartheta\\
&\times u^{\frac{N}{2}+1}e^{-2u}\left[1+O\!\left(u^{-1/2}\right)\right].
\end{aligned}
\label{eq:maxima_alpha0}
\end{equation}
where $\boldsymbol{\Lambda}(\vartheta)$ is the $\vartheta$-dependent effective second order spectral moment in Lemma~\ref{lem:speckle_size}, Eq.~\eqref{eq:two_beam_effective_moment_alpha0} and projected onto the corresponding plane using Eq.~\eqref{eq:Lambda_projection} when $N=2$.
\begin{proof}
See Appendix~\ref{sec:Proof_of_Theorem}.
\end{proof}
\end{theorem}

For completeness, the $\vartheta$ integral in Eq.~\eqref{eq:maxima_alpha0} for $N=3$ can be evaluated in closed form as
\[
\int_0^{\frac{\pi}{2}}\sin(2\vartheta)\sqrt{\det\!\boldsymbol{\Lambda}(\vartheta)}\,d\vartheta = \frac{\sqrt{\det\!\mathring{\mathbf{m}}^{(2)}}}{2}\left[1+\frac{\cos^{-1}\zeta}{\zeta\sqrt{1-\zeta^2}}\right],
\]
with $\zeta:=\det(2\,\mathring{\mathbf{m}}^{(2)})^{\frac{1}{2}}/\det(\mathring{\mathbf{m}}_1^{(2)}+\mathring{\mathbf{m}}_2^{(2)})^{\frac{1}{2}}$.

The asymptotic expression in Eq.~\eqref{eq:maxima_0alpha1}, valid for $0<\alpha\leq1$, is nonuniform as $\alpha\to0$ and does not reduce directly to the $\alpha=0$ result in Eq.~\eqref{eq:maxima_alpha0}. To connect these two limits, we therefore require an asymptotic expression for the transition regime $u\alpha=O(1)$,
\begin{equation}
\begin{aligned}
\left\langle M_\Omega^+(u)\right\rangle &={} \frac{2^{\frac{N}{2}+1}|\Omega|}{\pi^{N/2}}u^{\frac{N}{2}+1}e^{-2u}\int_0^{\pi/2}\sin(2\vartheta)\\
&\times\sqrt{\det\!\left(\sin^2\!\vartheta\,\mathring{\mathbf{m}}_1^{(2)}+\cos^2\!\vartheta\,\mathring{\mathbf{m}}_2^{(2)}\right)}\\
&\times\mathcal{I}_0\!\left(2\alpha u\sin(2\vartheta)\right)\left[1+O(u^{-1/2})\right]\,d\vartheta,
\end{aligned}
\label{eq:maxima_transition}
\end{equation}
where $\mathcal{I}_0$ denotes the modified Bessel function of the first kind of order zero.

In the left panel of Fig.~\ref{fig:speckle_abundance_results} we compare Eq.~\eqref{eq:maxima_0alpha1} from Theorem~\ref{the:speckle_abundance} with Monte Carlo simulations~\footnote{The number of local maxima in each realization is an integer, but the Monte Carlo simulation represent an average over 200 realizations and therefore may take fractional values.}, which shows convergence toward the asymptotic prediction as the intensity threshold increases. Fig.~\ref{fig:speckle_abundance_results} also illustrates the dependence on polarization overlap and crossing angle. Varying $\alpha$ changes the exponential decay rate of the speckle abundance, while varying $\gamma$ leaves the $u^{N/2}e^{-u}$ dependence unchanged and acts only through the geometric factor $\det(\boldsymbol{\Lambda})^{1/2}$.

\section{\label{sec:Quad_beams_SBS}Quad beams and SBS}
\begin{figure}[b]
\includegraphics[width=\linewidth]{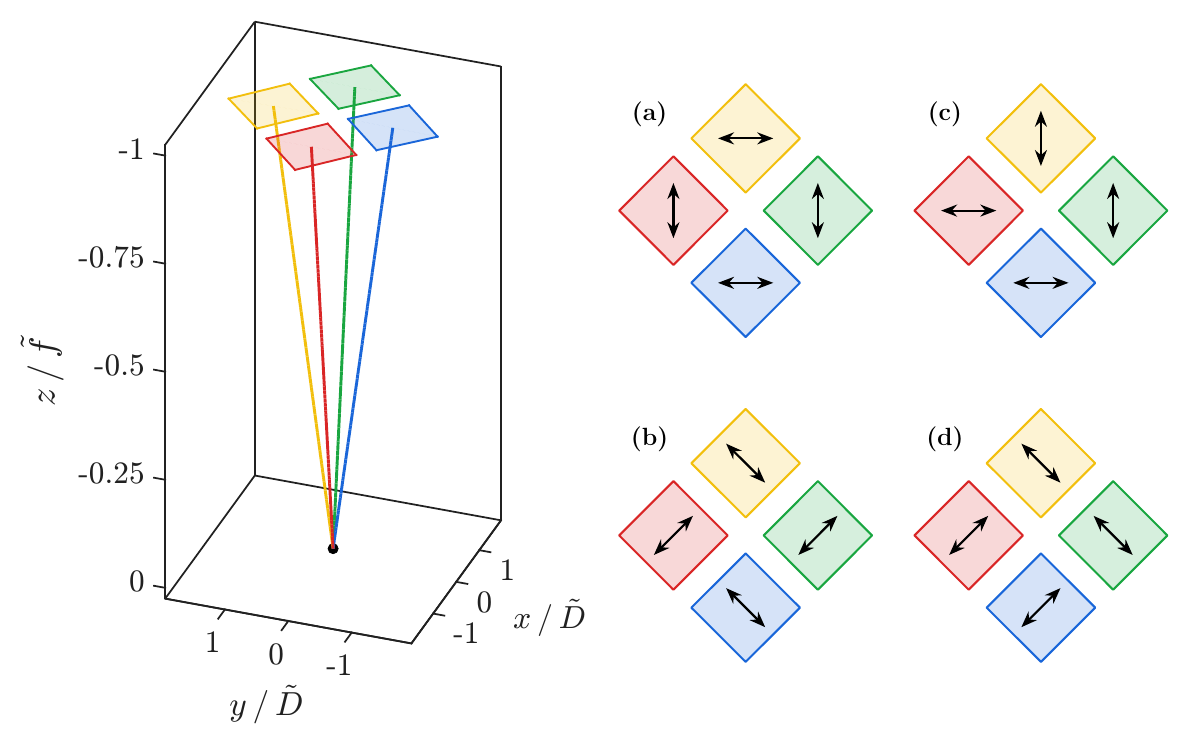}
\caption{Geometry and polarization configurations of four square aperture beams arranged in a quad. Left: four beams converge toward a common focal point, with colors distinguishing the individual beams. Right: polarization directions (black arrows) for the checkerboard \{(a),(b)\} and up-down \{(c),(d)\} configurations, with polarizations oriented either diagonally \{(a),(c)\} or along the aperture edges \{(b),(d)\}.}
\label{fig:quad_crossing_beams}
\end{figure}
To mitigate laser-plasma instabilities, laser facilities such as NIF apply polarization smoothing by overlapping the focal spots of four square aperture beams arranged in a quad, where two of the beams have identical linear polarization direction and the other two polarized orthogonally. The polarizations are aligned parallel to the aperture edges, with identically polarized beams occupying either "checkerboard" or "up-down" positions within the quad, as illustrated in Fig.~\ref{fig:quad_crossing_beams} (b) and (d), respectively~\cite{michel2023introduction}.

At finite crossing angles, the edge aligned polarizations are no longer exactly parallel or orthogonal due to the different beam propagation directions. For the checkerboard configuration, the two beam pairs propagate in mutually orthogonal planes, allowing polarizations along the diagonals to preserve these relationships exactly, Fig.~\ref{fig:quad_crossing_beams} (a). Since the edge aligned and diagonally aligned polarization choices are equivalent to leading order for small $\gamma$, we adopt the latter, which allows us to apply our two beam results from Section~\ref{sec:Two_Crossing_Beams} to each polarization pair. Unfortunately, the up-down configuration in Fig.~\ref{fig:quad_crossing_beams} (c) admits no choice of polarization directions that preserves these relationships at finite crossing angles, and therefore would require a separate calculation based on a more general treatment of crossing speckle beams.

\subsection{\label{sec:Vector_potential_intensity_4}Vector potential and intensity}
We consider the checkerboard configuration with diagonal polarizations illustrated in Fig.~\ref{fig:quad_crossing_beams} (a). We label the four beams such that the opposite pairs $(1,3)$ and $(2,4)$ have polarization directions aligned with the global $\hat{\mathbf{y}}$ and $\hat{\mathbf{x}}$ axes, respectively. For a pair $(j,k)$, we define the scalar pair field
\begin{equation}
\mathcal{A}_{j,k}:=\mathring{\mathcal{A}}_j e^{i\psi(z_j,t)}+\mathring{\mathcal{A}}_k e^{i\psi(z_k,t)},
\label{eq:4_pair_field}
\end{equation}
where $\psi$ is defined in Eq.~\eqref{eq:oscillating_phase_term}, so that the total vector potential can be written as
\begin{equation}
\mathbf{A}=\mathcal{A}_{1,3}\, \hat{\mathbf{y}} + \mathcal{A}_{2,4}\, \hat{\mathbf{x}}.
\label{eq:4_vector_potential}
\end{equation}
Since the individual beam envelopes are independent complex Gaussian fields~\cite{min2026statistical}, the two pairwise sums in Eq.~\eqref{eq:4_pair_field} are also independent complex Gaussian fields, each with distribution $\mathcal{CN}(0,2\sigma^2)$. Therefore, the normalized intensity $U:=I/\langle I \rangle$ given by
\begin{equation}
U=\frac{1}{4\sigma^2}\left(|\mathcal{A}_{1,3}|^2+|\mathcal{A}_{2,4}|^2\right),
\label{eq:4_normalized_intensity}
\end{equation}
has distribution
\begin{equation}
U\sim\frac{1}{4}\chi_4^2, \qquad \text{with PDF} \qquad f_U(u)=4u e^{-2u},
\label{eq:4_intensity_pdf}
\end{equation}
for $u\geq0$.

\subsection{\label{sec:Speckles_4}Speckles}
The propagation direction of a beam $j$ is
\begin{equation}
\hat{\mathbf{z}}_j=\sin\gamma\cos\phi_j\,\hat{\mathbf{x}}+\sin\gamma\sin\phi_j\,\hat{\mathbf{y}}+\cos\gamma\,\hat{\mathbf{z}},
\label{eq:4_beam_axes}
\end{equation}
where $\phi_j=(j-1)\frac{\pi}{2}$ and $\gamma$ is the angle of the beams propagation direction with respect to the global $\hat{\mathbf{z}}$ axis. Each pair field $\mathcal{A}_{j,k}$ defined in Eq.~\eqref{eq:4_pair_field} consists of two identically polarized beams, thus by Lemma~\ref{lem:speckle_size}, Eq.~\eqref{eq:2_effective_second_order_spectral_moment} with $\alpha=1$, its effective second order spectral moment is
\begin{equation}
\mathring{\mathbf{m}}_{j,k}^{(2)}=\frac{1}{2}\left[\mathring{\mathbf{m}}_j^{(2)}+\mathring{\mathbf{m}}_k^{(2)}\right]+\frac{\kappa_0^2}{4}\left[\hat{\mathbf{z}}_j-\hat{\mathbf{z}}_k\right]\left[\hat{\mathbf{z}}_j-\hat{\mathbf{z}}_k\right]^\top,
\label{eq:4_pair_second_spectral_moment}
\end{equation}
where $\hat{\mathbf{z}}_1-\hat{\mathbf{z}}_3 = 2\sin\!\gamma\,\hat{\mathbf{x}}$, $\hat{\mathbf{z}}_2-\hat{\mathbf{z}}_4 = 2\sin\!\gamma\,\hat{\mathbf{y}}$, and $\mathring{\mathbf{m}}_j^{(2)}$ is defined in Eq.~\eqref{eq:spectral_moment_rotation}. We then apply Lemma~\ref{lem:speckle_size} once more using Eq.~\eqref{eq:two_beam_effective_moment_alpha0} with $\alpha=0$, so that the effective second order spectral moment for the quad beams is
\begin{equation}
\boldsymbol{\Lambda}(\vartheta)=\sin^2\vartheta\,\mathring{\mathbf{m}}_{1,3}^{(2)}+\cos^2\vartheta\,\mathring{\mathbf{m}}_{2,4}^{(2)},
\label{eq:4_effective_second_order_spectral_moment}
\end{equation}
for parametrization variable $\vartheta$ defined in Lemma~\ref{lem:speckle_size}. The effective speckle length along the global $\hat{\mathbf{z}}$ direction, $\ell_z = \left[\hat{\mathbf{z}}^T\boldsymbol{\Lambda}(\vartheta)\hat{\mathbf{z}}\right]^{-1/2}$, follows from Eq.~\eqref{eq:speckle_lengths} which by direct calculation is
\begin{equation}
\ell_z(\gamma)=\left(\lambda_\perp\sin^2\gamma+\lambda_\parallel\cos^2\gamma\right)^{-1/2},
\label{eq:4_effective_speckle_length_z_final}
\end{equation}
where $\lambda_\perp$ and $\lambda_\parallel$ are given for a square flat-top aperture in Eq.~\eqref{eq:lambda_eigenvalues_square}.
\begin{corollary}[Number of Speckles for Quad Beams]\label{cor:4_speckle_abundance}
For $N=2,3$, let $U:\mathbb{R}^N\to[0,\infty)$ denote the normalized intensity field defined in Eq.~\eqref{eq:4_normalized_intensity}, and restricted to a plane when $N=2$, and let $M_\Omega^+(u)$ denote the number of speckles with centres $\mathbf{x}_*\in\Omega$ satisfying $U(\mathbf{x}_*)>u$. Then as $u\to\infty$,
\begin{equation}
\begin{aligned}
\left\langle M_\Omega^+(u)\right\rangle ={}& \frac{2^{\frac{N}{2}+1}|\Omega|}{\pi^{N/2}} \int_0^{\frac{\pi}{2}}\sin(2\vartheta)\sqrt{\det\!\boldsymbol{\Lambda}(\vartheta)}\,d\vartheta\\
&\times u^{\frac{N}{2}+1}e^{-2u} \left[1+O\!\left(u^{-1/2}\right)\right],
\end{aligned}
\label{eq:4_speckle_abundance}
\end{equation}
where $\boldsymbol{\Lambda}(\vartheta)$ is given by Eq.~\eqref{eq:4_effective_second_order_spectral_moment}, and projected using Eq.~\eqref{eq:Lambda_projection} when $N=2$.
\begin{proof}
At any fixed spatial point, the underlying fields, $\mathcal{A}_{1,3}$ and $\mathcal{A}_{2,4}$, of $U$ defined in Eq.~\eqref{eq:4_normalized_intensity} are independent proper complex Gaussian random variables, each distributed as $\mathcal{CN}(0,2\sigma^2)$. This implies that $U$ can be written in the form of Eq.~\eqref{eq:crossing_beam_intensity} with $\alpha=0$ by replacing the two single-beam fields with $\mathcal{A}_{1,3}$ and $\mathcal{A}_{2,4}$ and their second-order spectral moments by $\mathring{\mathbf{m}}_{1,3}^{(2)}$ and $\mathring{\mathbf{m}}_{2,4}^{(2)}$. Thus, the $\alpha=0$ result in Eq.~\eqref{eq:maxima_alpha0} of Theorem~\ref{the:speckle_abundance} applies directly with the $\vartheta$-dependent effective second-order spectral moment $\boldsymbol{\Lambda}(\vartheta)$ defined in Eq.~\eqref{eq:4_effective_second_order_spectral_moment}.
\end{proof}
\end{corollary}
Corollary~\ref{cor:4_speckle_abundance} also illustrates the statistical benefit of polarization smoothing. For a single beam, the threshold abundance obtained in Ref.~\cite{min2026statistical} scaled as $u^{N/2}e^{-u}$, whereas the polarization smoothed quad field scales as $u^{N/2+1}e^{-2u}$, so the expected number of speckles above a level set is strongly reduced relative to the average intensity.

\subsection{\label{sec:Reflectivity_saturation}Reflectivity saturation}
We now consider the SBS reflectivity from saturated speckles. The speckle field is treated as a prescribed intensity imprint on which the local SBS response developes. The intense speckle saturation model from Rose~\cite{rose1994laser} states that speckles with peak intensities exceeding a threshold $u_{\mathrm{th}}$ are taken to be locally saturated and reflects the power they contain. While this model provides only a limited description of SBS, it can be useful for determining the contribution of saturated speckles to the mean SBS reflectivity as the separation of the quad beams are varied.

To find the mean reflectivity, we let $\mathcal{M}_\Omega^+(u,\vartheta)\,d\vartheta$ denote the expected number of speckles in $\Omega$ with maxima exceeding $u$, parametrized in $[\vartheta,\vartheta+d\vartheta]$. If $P(u,\vartheta)$ denotes the fraction of incident beam power carried by such a speckle, then
\begin{equation}
\langle R\rangle=-\int_0^{\pi/2}\int_{u_{\mathrm{th}}}^\infty P(u,\vartheta)\frac{\partial \mathcal{M}_\Omega^+(u,\vartheta)}{\partial u}\,du\,d\vartheta.
\label{eq:reflectivity}
\end{equation}

For a single beam, the natural backscatter direction is antiparallel to its propagation axis, along which the speckles are most elongated. However, a quad configuration consists of four distinct beam propagation axes, so we consider backscatter antiparallel to the central quad axis $-\hat{\mathbf{z}}$, for which all four beams are displaced by the same angle $\gamma$ from exact backscatter. 

The power carried by a speckle is determined by its cross section normal to the common backward direction. So for the effective second order spectral moment found in Eq.~\eqref{eq:4_effective_second_order_spectral_moment}, its transverse cross section $\boldsymbol{\Lambda}_\perp(\vartheta)$, is obtained by projecting $\boldsymbol{\Lambda}(\vartheta)$ onto the global $(x,y)$ plane, so that $\boldsymbol{\Lambda}_\perp(\vartheta) = (\hat{\mathbf{x}}\ \hat{\mathbf{y}})^\top\boldsymbol{\Lambda}(\vartheta)(\hat{\mathbf{x}}\ \hat{\mathbf{y}})$. We write the spatial domain as $\Omega=\Omega_\perp\times\Omega_z$, where $\Omega_\perp$ is the transverse domain and $\Omega_z$ is the longitudinal domain along the global $z$-axis, with $|\Omega_z|$ representing the interaction length. Then the fraction of the incident power carried by a speckle of normalized intensity $u$ is then
\begin{equation}
P(u;\vartheta,\gamma)=\frac{\pi u}{|\Omega_\perp|\sqrt{\det\boldsymbol{\Lambda}_\perp}}.
\label{eq:quad_speckle_power}
\end{equation}
Substituting Eq.~\eqref{eq:quad_speckle_power} and the $\vartheta$-resolved speckle abundance underlying Corollary~\ref{cor:4_speckle_abundance} for $N=3$ into Eq.~\eqref{eq:reflectivity} gives
\[
\begin{aligned}
\left\langle R\right\rangle &= \frac{2^{5/2}|\Omega_z|}{\sqrt{\pi}}\int_0^{\pi/2}\sin(2\vartheta)\sqrt{\frac{\det\boldsymbol{\Lambda}(\vartheta)}{\det\boldsymbol{\Lambda}_\perp(\vartheta)}}\,d\vartheta\\
&\times\int_{u_{\mathrm{th}}}^{\infty} \left(2u^{\frac{7}{2}}-\frac{5}{2}u^{\frac{5}{2}}\right)e^{-2u}\left[1+O\!\left(u^{-\frac{1}{2}}\right)\right]\,du.
\end{aligned}
\]
From Refs.~\cite{michel2023introduction,rose1994laser} we define $u_{\mathrm{th}}$ as the normalized speckle intensity for which the local reflectivity reaches unity, taking the larger solution of
\begin{equation}
R_0\exp\left[G(\gamma)u_{\mathrm{th}}(\gamma)\right]=u_{\mathrm{th}}(\gamma),
\label{eq:saturation_threshold}
\end{equation}
where $R_0\ll 1$ is the seed reflectivity and $G(\gamma)$ is the speckle gain at normalized intensity, which for scattering along $-\hat{\mathbf{z}}$, we define as
\begin{equation}
G(\gamma)=g\langle I\rangle\ell_z(\gamma),
\label{eq:quad_speckle_gain}
\end{equation}
where $g$ is the SBS gain coefficient and assumed to be independent of $\gamma$~\footnote{While this is an assumption of our model, angular independence of steady-state spatial growth rate had also been established for SBS in homogeneous plasma using a linear model~\cite{giacone1995angular}.}. So while $u_{\mathrm{th}}$ varies with $\gamma$, we still require $u_{\mathrm{th}}\gg 1$, which allows us to use Laplace's method for endpoint asymptotics to solve the $u$ integral
\begin{equation}
\langle R\rangle = \frac{2^{\frac{5}{2}}|\Omega_z|}{\sqrt{\pi}\,\ell_z(\gamma)}\,u_{\mathrm{th}}^{7/2}(\gamma)e^{-2u_{\mathrm{th}}(\gamma)}\left[1+O\!\left(u_{\mathrm{th}}^{-1/2}\right)\right],
\label{eq:quad_reflectivity}
\end{equation}
where $\ell_z(\gamma)$ is given by Eq.~\eqref{eq:4_effective_speckle_length_z_final}, and the $\vartheta$ integral was calculated directly as
\[
\int_0^{\pi/2}\sin(2\vartheta)\sqrt{\frac{\det\boldsymbol{\Lambda}(\vartheta)}{\det\boldsymbol{\Lambda}_\perp(\vartheta)}}\,d\vartheta = \frac{1}{\ell_z(\gamma)}.
\]

To determine whether the expected SBS reflectivity increases or decreases with quad beam separation $\gamma$, we examine the sign of its derivative with respect to $\gamma$. This is most conveniently accomplished by taking the logarithmic derivative of Eq.~\eqref{eq:quad_reflectivity}, which to leading order gives
\begin{equation}
\partial_\gamma\ln\langle R\rangle = -\frac{\partial_\gamma\ell_z}{\ell_z}+\left(\frac{7}{2u_{\mathrm{th}}}-2\right)\partial_\gamma u_{\mathrm{th}},
\label{eq:log_derivative_reflectivity}
\end{equation}
since both the speckle length $\ell_z$ and the threshold intensity $u_{\mathrm{th}}$ depends on $\gamma$ which is given implicitly in Eq.~\eqref{eq:saturation_threshold}. To find its derivative with respect to $\gamma$, we can implicitly take the logarithmic derivative of Eq.~\eqref{eq:saturation_threshold} to get
\begin{equation}
\partial_\gamma u_{\mathrm{th}}(\gamma)=-\frac{\partial_\gamma G(\gamma)\, u_{\mathrm{th}}^2(\gamma)}{G(\gamma) u_{\mathrm{th}}(\gamma)-1}.
\label{eq:log_derivative_saturation_threshold}
\end{equation}
Under the assumption given for Eq.~\eqref{eq:quad_speckle_gain}, the angular dependence of the speckle gain is determined entirely by the longitudinal speckle length, and since $\lambda_\perp>\lambda_\parallel$, logarithmic differentiating Eq.~\eqref{eq:quad_speckle_gain} with respect to $\gamma$ gives
\begin{equation}
\frac{\partial_\gamma G(\gamma)}{G(\gamma)}=\frac{\partial_\gamma\ell_z(\gamma)}{\ell_z(\gamma)}=-\frac{(\lambda_\perp-\lambda_\parallel)\sin\gamma\cos\gamma}{\lambda_\perp\sin^2\gamma+\lambda_\parallel\cos^2\gamma}<0,
\label{eq:gain_derivative_fraction}
\end{equation}
for $0<\gamma<\pi/2$. Substituting this into Eq.~\eqref{eq:log_derivative_saturation_threshold} gives us 
\[
\partial_\gamma u_{\mathrm{th}} =-\frac{\partial_\gamma\ell_z}{\ell_z} \frac{G u_{\mathrm{th}}^2}{G u_{\mathrm{th}} - 1} > -\frac{\partial_\gamma \ell_z}{\ell_z}u_{\mathrm{th}},
\]
where the new inequality is algebraically implied, since, by Eq.~\eqref{eq:saturation_threshold}, $G(\gamma)u_{\mathrm{th}}(\gamma)>1$. The inequality from Eq.~\eqref{eq:gain_derivative_fraction} also implies that $u_{\mathrm{th}}^\prime(\gamma)>0$, so increasing the beam crossing angle raises the normalized intensity required for a speckle to reach saturation, which is intuitive since the effective speckle length decreases as $\gamma$ increases. So combining this with the logarithmic derivative of the mean reflectivity from Eq.~\eqref{eq:log_derivative_reflectivity} gives
\begin{equation}
\frac{d}{d\gamma}\ln\langle R\rangle < -\frac{\ell_z'(\gamma)}{\ell_z(\gamma)}\left(\frac{9}{2}-2u_{\mathrm{th}}(\gamma)\right)<0,
\label{eq:quad_reflectivity_derivative_bound}
\end{equation}
provided $u_{\mathrm{th}}>9/4$, a condition already satisfied in the high-threshold regime $u_{\mathrm{th}}\gg1$ required for Eq.~\eqref{eq:quad_reflectivity}. Since $\langle R\rangle>0$, the mean reflectivity decreases monotonically with increasing crossing angle in the leading order asymptotic approximation.

\begin{figure}
\includegraphics[width=\linewidth]{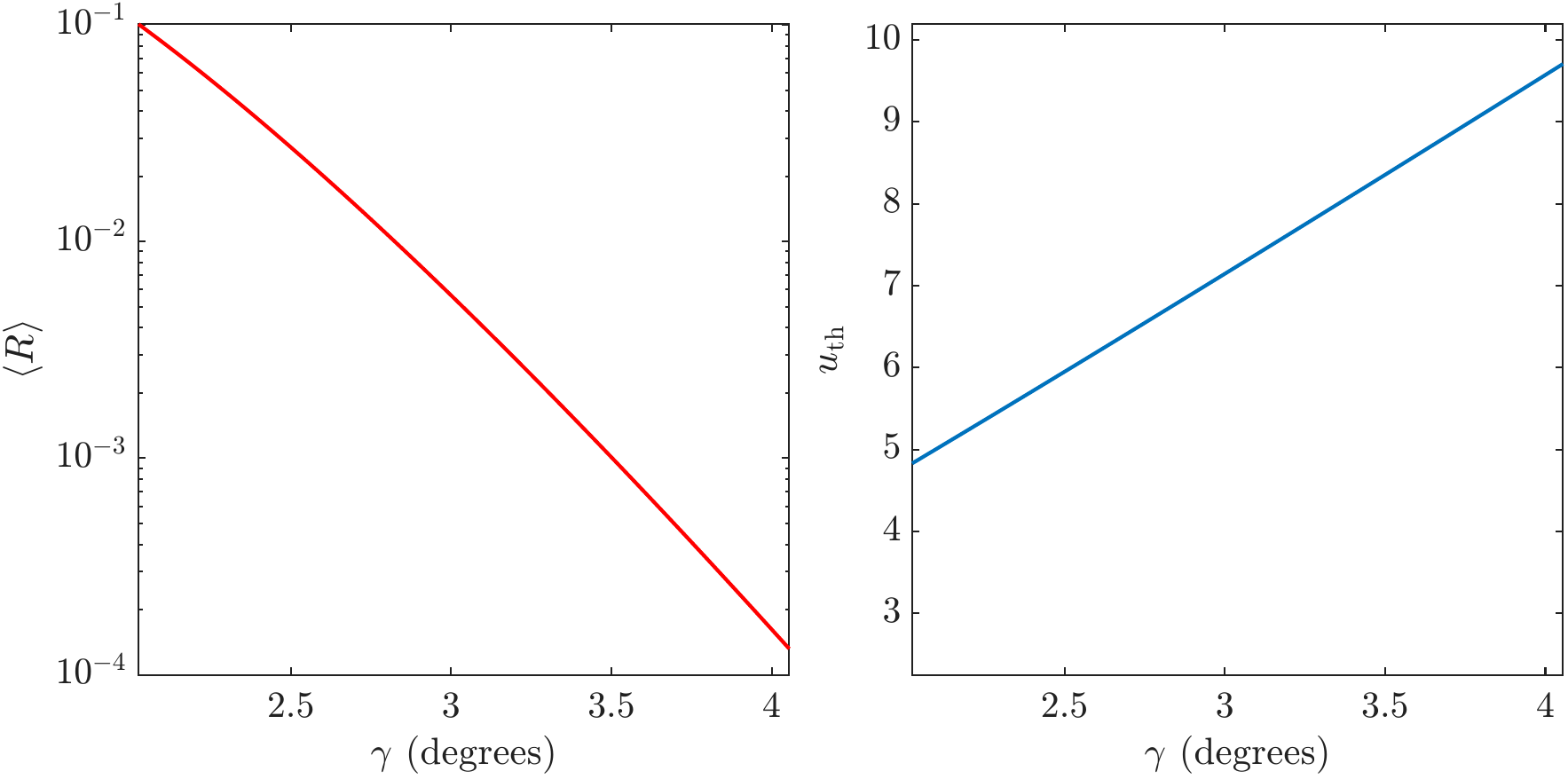}
\caption{Left: Mean SBS reflectivity from saturated speckles as a function of the crossing angle $\gamma$ for the checkerboard polarization configuration Eq.~\eqref{eq:quad_reflectivity} where $u_{\mathrm{th}}$ is given implicitly in Eq.~\eqref{eq:saturation_threshold}. Right: Intensity threshold $u_{\mathrm{th}}$ as a function of $\gamma$ Eq.~\eqref{eq:saturation_threshold}. Parameters used for both graphs: $\lambda_0=351\,\mathrm{nm}$, $f_\#=20$, $\langle I\rangle=4\times10^{15}\,\mathrm{W/cm^2}$, $g=1.2\times10^{-15}\,\mathrm{m/W}$, $R_0=10^{-10}$, $|\Omega_z|=6.37\times10^{-4}\,\mathrm{m}$ and $\gamma\in[2.03^\circ,4.05^\circ]$.}
\label{fig:quad_reflectivity}
\end{figure}

In Fig.~\ref{fig:quad_reflectivity} (left) we plot Eq.~\eqref{eq:quad_reflectivity} for a relevant range of angular separation $\gamma\in[\gamma_{\min},\gamma_{\max}]$, showing that $\langle R\rangle$ decreases with increasing $\gamma$. The lower bound $\gamma_{\min}:=\arctan(1/(\sqrt{2}f_\#))$ corresponds to when the four square apertures just touch along their edges, and the upper bound $\gamma_{\max}:=\arcsin(\sqrt{2}/f_\#)$ corresponds to the adjacent apertures being separated by approximately one aperture width. Here, $f_{\#}=\tilde{f}/\tilde{D}$ is the $f$ number of the beam. We had also set the seed reflectivity to $10^{-10}$ and set the SBS gain coefficient to $g=1.2\times10^{-15}\, \mathrm{m/W}$ so that $u_{\mathrm{th}}$ remains in a relevant regime, as shown in Fig.~\ref{fig:quad_reflectivity} (right).

\section{\label{sec:Extension_to_arbitrary_number_beams}Extension to an arbitrary number of beams}
\begin{figure*}
\raisebox{0.34\linewidth}{\makebox[0pt][r]{(a)\hspace{0.3em}}}%
\includegraphics[width=0.45\linewidth]{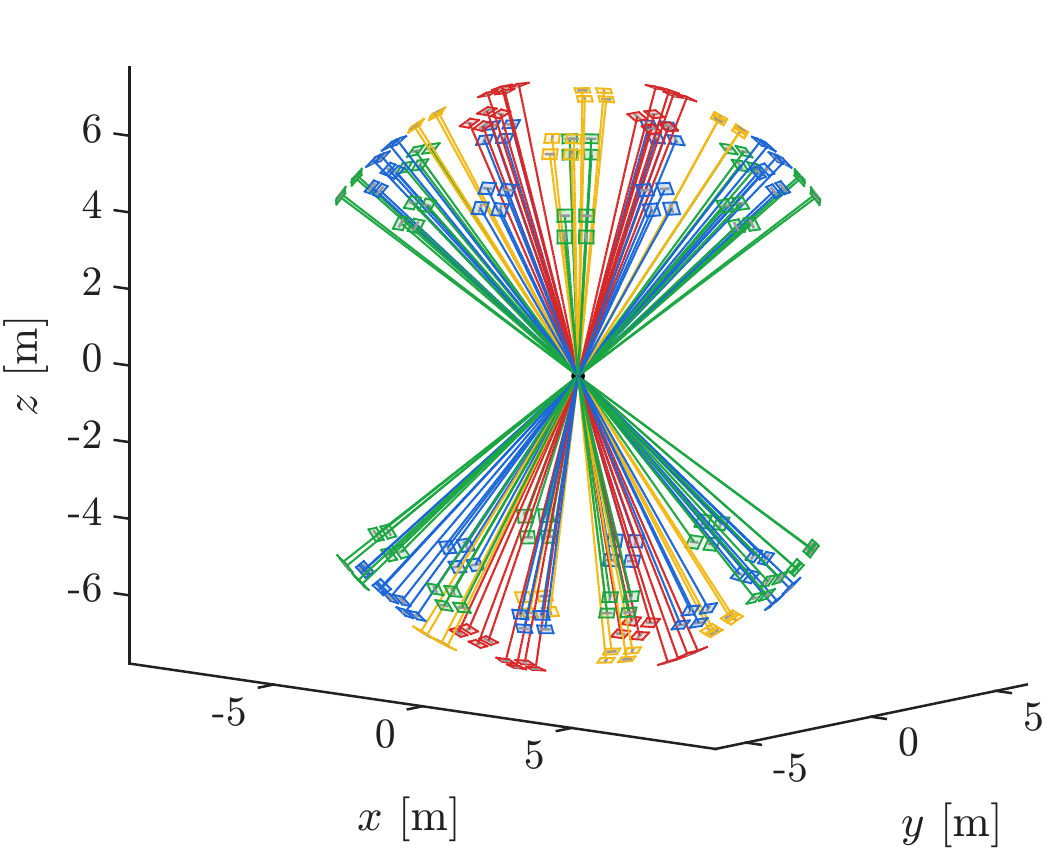}\hfill
\raisebox{0.34\linewidth}{\makebox[0pt][r]{(b)\hspace{0.3em}}}%
\includegraphics[width=0.45\linewidth]{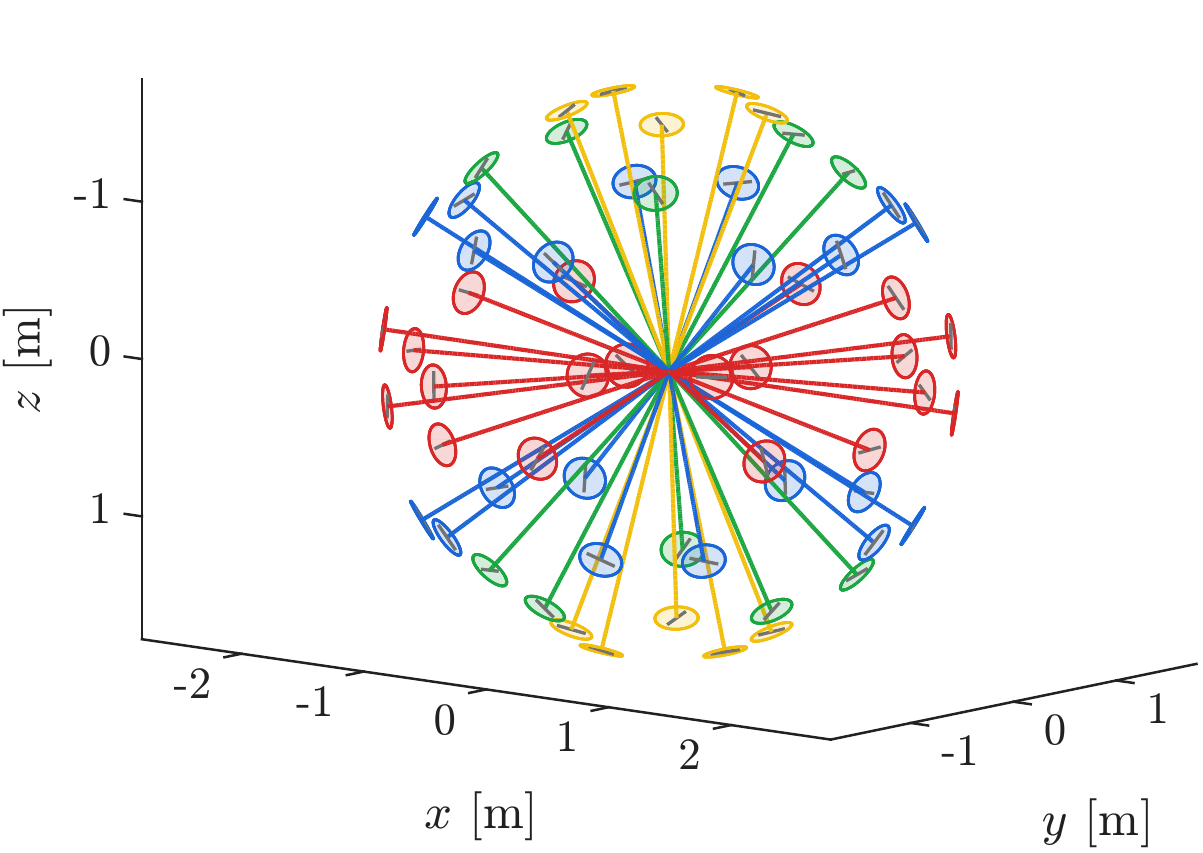}
\caption{Numerically constructed beam geometries used in the multi-beam simulations for (a) the 192-beam National Ignition Facility (NIF) and (b) the 60-beam OMEGA Laser System, assuming all beams are simultaneously active and their axes converge at the origin.}
\label{fig:M_crossing_beams}
\end{figure*}

Laser driven electron density fluctuations can produce spectral features distinct from those arising from thermal plasma fluctuations~\cite{min2026statistical}. Interference between overlapping speckle beams introduces additional spatial structure determined by their relative propagation directions and polarizations~\cite{le2011statistical}. We therefore extend the statistical analysis to an arbitrary number of beams in a coordinate free formulation, deriving the intensity correlations and the spectral density of the resulting electron density fluctuations for general crossing beam configurations, such as that illustrated in Fig.~\ref{fig:M_crossing_beams}.

\subsection{\label{sec:Vector_potential_and_intensity_M}Vector potential and intensity}
Consider $M$ independently generated paraxial speckled beams, each with a phase corrected envelope $\mathring{\mathcal{A}}_j\colon\mathbb{R}^3\to\mathbb{C}$ as described in Eq.~\eqref{eq:phase_corrected_envelope}. For each beam we consider a local coordinate system spanning the basis $\{\hat{\mathbf{x}}_j,\hat{\mathbf{y}}_j,\hat{\mathbf{z}}_j\}$ such that $\hat{\mathbf{y}}_j$ denotes the polarization direction and $\hat{\mathbf{z}}_j$ denotes the propagation direction of the beam. The total complex vector potential $\mathbf{A}\colon\mathbb{R}^4\to\mathbb{C}^3$ of all crossing beams is the superposition of each beam
\begin{equation}
\mathbf{A}(x,y,z,t) = \sum_{j=1}^{M} \mathring{\mathcal{A}}_j(x_j,y_j,z_j)\,e^{i\psi(z_j,t)} \hat{\mathbf{y}}_j,
\label{eq:M_beam_field}
\end{equation}
where $\psi$ is the oscillatory phase term defined in Eq.~\eqref{eq:oscillating_phase_term}. The scalar envelopes are mutually independent and the deterministic phases do not alter their distributions; $\mathring{\mathcal{A}}_j\sim\mathcal{CN}(0,\sigma^2)$ implies $\mathbf{A}\sim\mathcal{CN}(\mathbf{0},\boldsymbol{\Sigma})$, where the one-point covariance matrix is defined as
\begin{equation}
\boldsymbol{\Sigma} = \sigma^2\sum_{j=1}^{M}\hat{\mathbf{y}}_j\otimes\hat{\mathbf{y}}_j.
\end{equation}
The covariance matrix $\boldsymbol{\Sigma}$ has $r=\operatorname{rank}(\boldsymbol{\Sigma})\leq3$ nonzero eigenvalues that can be expressed as $M\sigma^2\,w_q$ for $q=1,\ldots,r$. Since $\boldsymbol{\Sigma}$ is positive semidefinite and $\operatorname{tr}(\boldsymbol{\Sigma})=M\sigma^2$, the weights satisfy $w_q>0$ and $\sum_{q=1}^r w_q = 1$.

The intensity field is defined as Eq.~\eqref{eq:intensity_definition}, then its mean can be calculated as $\langle I\rangle=MI_0$, where $I_0$ is the mean intensity of a single beam defined in Eq.~\eqref{eq:mean_intensity}. The normalized intensity $U:=I/\langle I\rangle$ then follows a weighted sum of chi-squared distributions. To see this, consider the orthogonal matrix $\mathbf{Q}$ with columns that are the eigenvectors of $\boldsymbol{\Sigma}$. Since $\mathbf{Q}$ diagonalizes $\boldsymbol{\Sigma}$, the components of $\mathbf{Q}^\top\mathbf{A}$ are independent complex Gaussian random variables. Furthermore, since $\mathbf{Q}$ is orthogonal, $\lvert\mathbf{Q}^\top\mathbf{A}\rvert^2 = \lvert\mathbf{A}\rvert^2$, implying
\begin{equation}
U = \frac{|\mathbf{A}|^2}{M\sigma^2} \sim \frac{1}{2}\sum_{q=1}^{r}w_q\chi_{2,q}^{2}.
\label{eq:M_beam_weighted_chi}
\end{equation}
The probability density function of this weighted sum is derived in Appendix~\ref{sec:Probability_density_function}, which for distinct nonzero weights is
\begin{equation}
f_U(u)=\sum_{q=1}^{r}\left[\prod_{\substack{p=1\\p\neq q}}^{r}\frac{w_q}{w_q-w_p}\right]\frac{1}{w_q}\exp\left(-\frac{u}{w_q}\right),
\label{eq:M_beam_intensity_pdf}
\end{equation}
for $u\geq0$. If two or more weights are identical, the probability density can be obtained by taking the appropriate limit. Equation~\eqref{eq:M_beam_intensity_pdf} shows that, regardless of the number of incident beams, the one-point intensity statistics are determined by at most three normalized nonzero eigenvalues of the covariance matrix $\boldsymbol{\Sigma}$.

\begin{figure*}
\raisebox{0.34\linewidth}{\makebox[0pt][r]{(a)\hspace{0.3em}}}%
\includegraphics[width=0.45\linewidth]{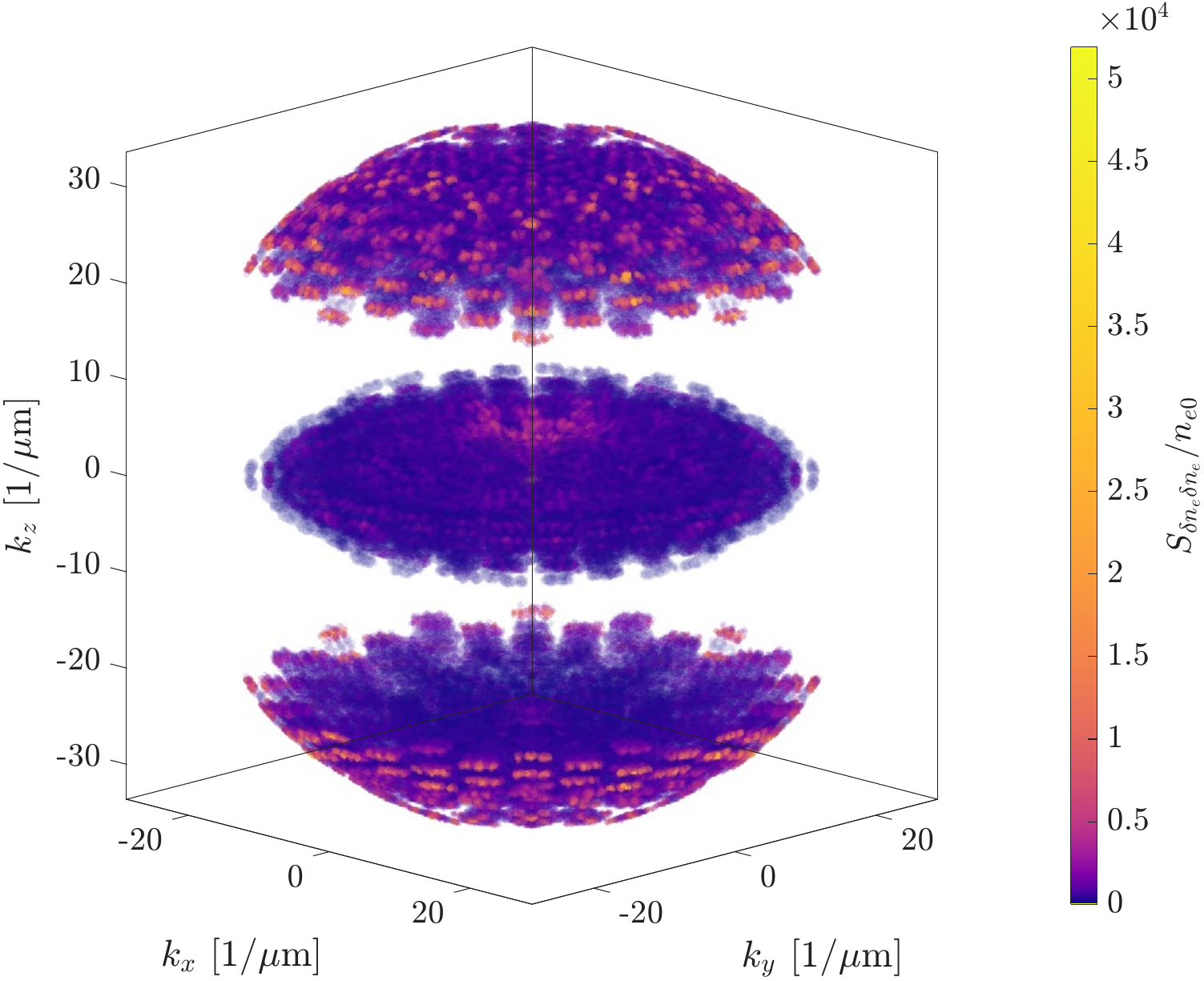}\hfill
\raisebox{0.34\linewidth}{\makebox[0pt][r]{(b)\hspace{0.3em}}}%
\includegraphics[width=0.45\linewidth]{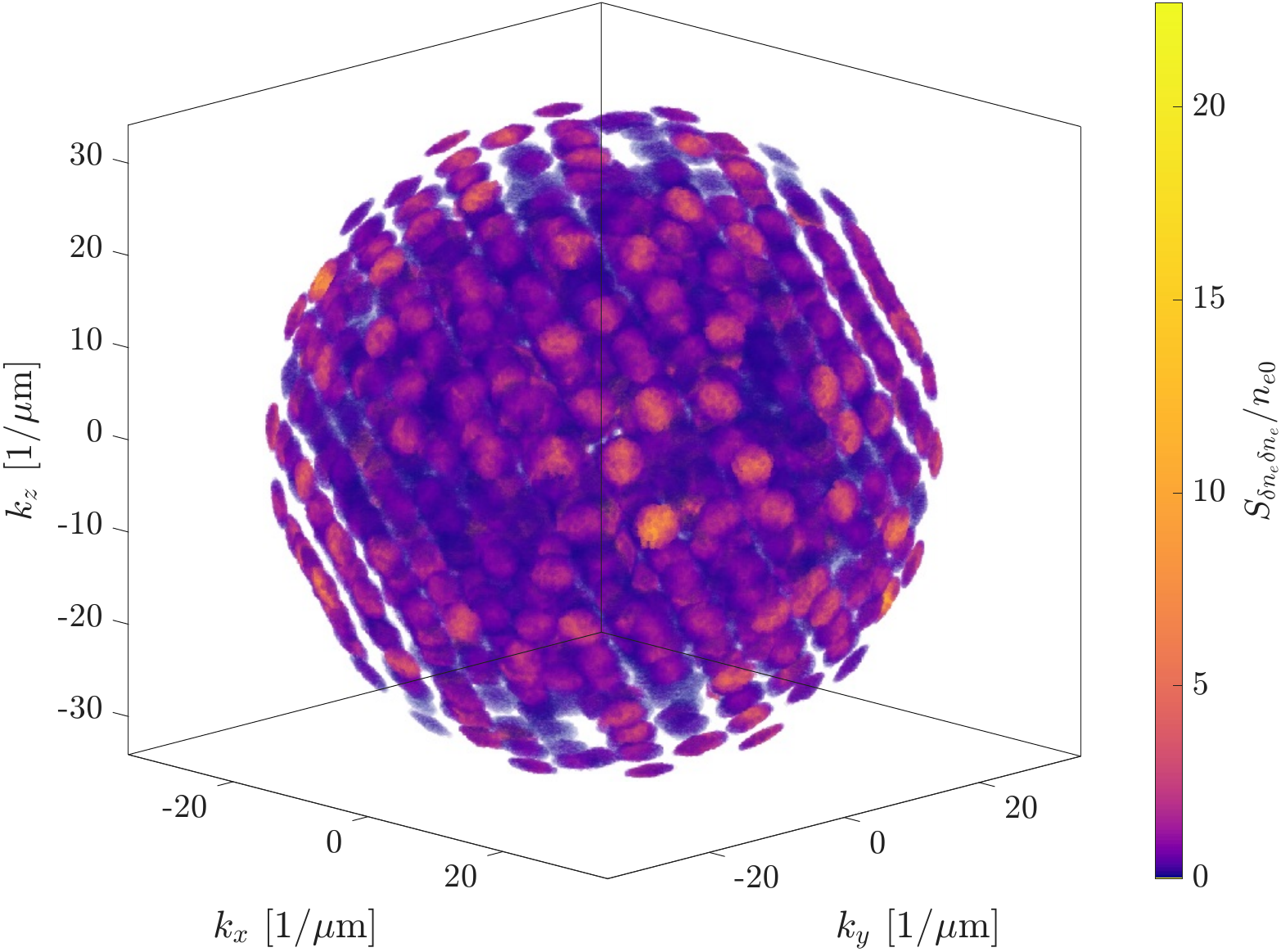}
\vspace{0.8em}
\raisebox{0.40\linewidth}{\makebox[0pt][r]{(c)\hspace{0.3em}}}%
\includegraphics[width=0.45\linewidth]{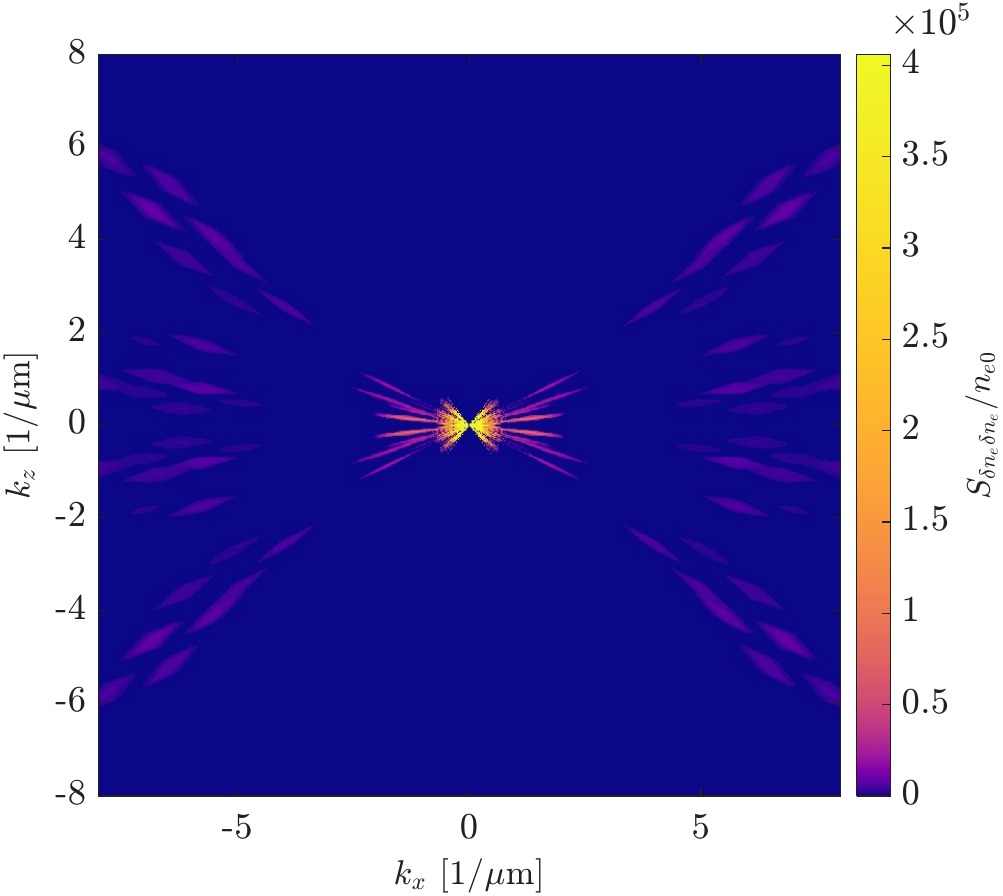}\hfill
\raisebox{0.40\linewidth}{\makebox[0pt][r]{(d)\hspace{0.3em}}}%
\includegraphics[width=0.45\linewidth]{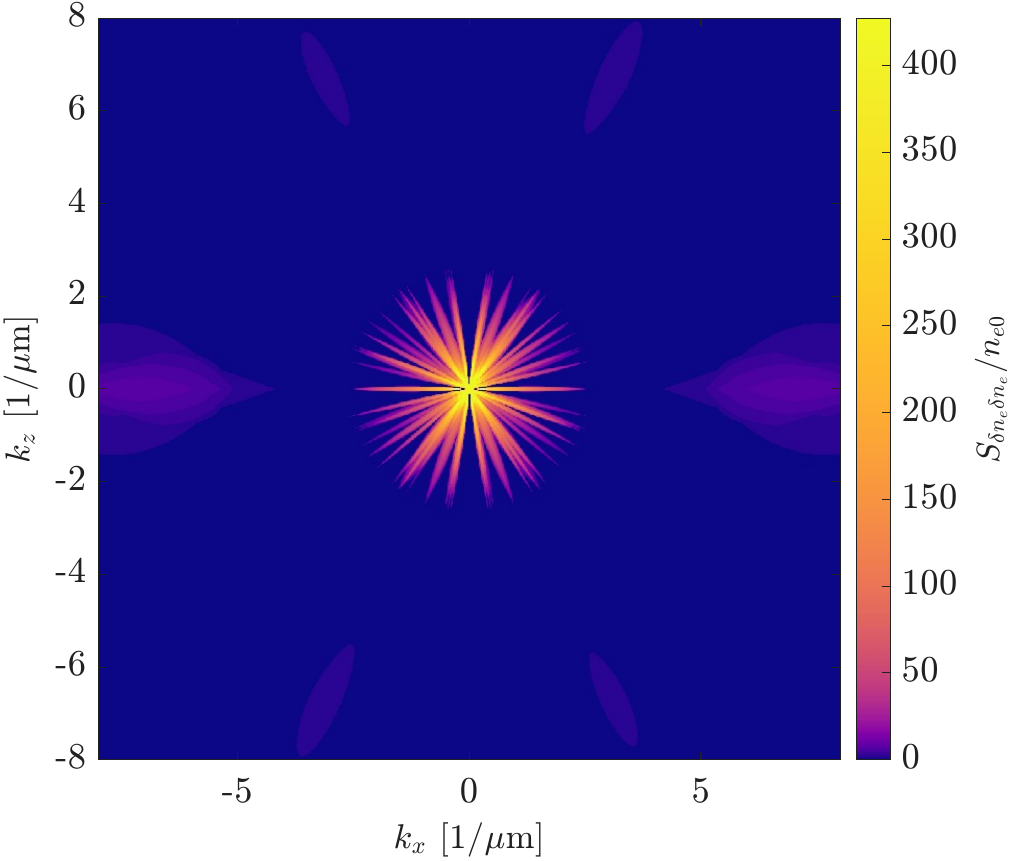}
\caption{The spectral density of electron density fluctuations, Eq.~\eqref{eq:density_fluctuations_expand}, for the NIF (a,c) and OMEGA (b,d) configurations in Fig.~\ref{fig:M_crossing_beams}. Panels (a,b) show the three-dimensional spectra with the individual-beam region excluded, while (c,d) show the $k_y=0$ plane, including both individual-beam speckle and crossing-beam interference contributions. The colour scales in (c,d) are capped at the 99.5th percentile of positive values. For (a,c), $f_\#=20.7$, $\langle I\rangle=10^{16}\,\mathrm{W/cm^2}$, $n_{e0}=10^{20}\,\mathrm{cm}^{-3}$, $T_e=2600\,\mathrm{eV}$, and $Z=2$ (He); for (b,d), $f_\#=6.7$, $\langle I\rangle=10^{14}\,\mathrm{W/cm^2}$, $n_{e0}=2\times10^{19}\,\mathrm{cm}^{-3}$, $T_e=260\,\mathrm{eV}$, and $Z=5$ (N). Both use $\lambda_0=351\,\mathrm{nm}$ and $T_i=T_e/2$.}
\label{fig:Density_spectrum}
\end{figure*}

\subsection{\label{sec:Two_point_covariance_function_M}Two-point statistics}
For a statistically homogeneous and temporally stationary random field, the two-point covariance of the complex vector potential is defined as
\[
\mathbf{C}_{\mathbf{A}\mathbf{A}}(\mathbf{x},t) := \left\langle\mathbf{A}(\mathbf{x},t)\mathbf{A}(\mathbf{0},0)^\dagger\right\rangle,
\]
where $\mathbf{x}\in\mathbb{R}^3$ and $t\in\mathbb{R}$ denote the spatial and temporal separations, respectively, and $\mathbf{C}_{\mathbf{A}\mathbf{A}}(\mathbf{0},0)=\boldsymbol{\Sigma}$. Using Eq.~\eqref{eq:M_beam_field} and the statistical independence of the beams, we obtain
\begin{equation}
\mathbf{C}_{\mathbf{A}\mathbf{A}}(\mathbf{x},t) = \sum_{j=1}^{M}C_{\mathring{\mathcal{A}}_j\mathring{\mathcal{A}}_j}(x_j,y_j,z_j)e^{i\psi(z_j,t)}\hat{\mathbf{y}}_j\otimes\hat{\mathbf{y}}_j,
\label{eq:M_beam_covariance_A}
\end{equation}
where $C_{\mathring{\mathcal{A}}_j\mathring{\mathcal{A}}_j}$ is the spatial covariance of the phase corrected scalar envelope of beam $j$~\cite{min2026statistical}, and $(x_j,y_j,z_j)$ are obtained from the separation coordinates through the rotation matrix in Eq.~\eqref{eq:coordinate_rotation}. The two-point covariance function of the normalized intensity field is defined as
\[
C_{UU}(\mathbf{x},t) := \langle U(\mathbf{x},t)U(\mathbf{0},0)\rangle-1,
\]
and since $\mathbf{A}$ is a zero-mean proper complex Gaussian random field, by Wick's probability theorem
\begin{equation}
C_{UU}(\mathbf{x},t) = \frac{\operatorname{tr}\left[\mathbf{C}_{\mathbf{A}\mathbf{A}}(\mathbf{x},t)\mathbf{C}_{\mathbf{A}\mathbf{A}}(\mathbf{x},t)^\dagger\right]}{(\operatorname{tr}\boldsymbol{\Sigma})^2},
\label{eq:M_beam_covariance_U}
\end{equation}
where $(\operatorname{tr}\boldsymbol{\Sigma})^2 = M^2\sigma^4$. The (equal-time spatial) spectral density is defined as
\[
S_{UU}(\mathbf{k}) = \mathcal{F}\left[C_{UU}(\mathbf{x},0)\right],
\]
where $\mathcal{F}: L^2(\mathbb{R}^3) \to L^2(\mathbb{R}^3)$ denotes the Fourier transform such that $\mathbf{x}\mapsto\mathbf{k}$ under the non-unitary convention. Then using Eqs.~\eqref{eq:M_beam_covariance_A} and \eqref{eq:M_beam_covariance_U}, we get that
\begin{equation}
S_{UU}(\mathbf{k}) = \frac{1}{M^2}\sum_{m,n=1}^{M}\left\lvert\hat{\mathbf{y}}_m\cdot\hat{\mathbf{y}}_n\right\rvert^2 S_{mn}(\mathbf{k}),
\label{eq:M_beam_spectrum_U}
\end{equation}
where $S_{mn}$ is the normalized spectral density contribution associated with beam pair $(m,n)$, defined as
\begin{equation}
S_{mn}(\mathbf{k}) = \frac{1}{\sigma^4}\,\mathcal{F}\left[C_{\mathring{\mathcal{A}}_m\mathring{\mathcal{A}}_m}\overline{C_{\mathring{\mathcal{A}}_n\mathring{\mathcal{A}}_n}} e^{i\kappa_0(z_m-z_n)}\right],
\label{eq:M_beam_spectrum_pair}
\end{equation}
such that
\[
\int_{\mathbb{R}^3} S_{mn}(\mathbf{k}) \,d\mathbf{k} = (2\pi)^3.
\]
Eq.~\eqref{eq:M_beam_spectrum_U} separates into two sets of contributions: the individual beam spectrum ($m=n$) and the interference spectra associated with distinct beam pairs ($m\neq n$), so
\begin{equation}
\begin{aligned}
S_{UU}(\mathbf{k}) =&{} \frac{1}{M^2}\sum_{m=1}^{M}S_{mm}(\mathbf{k})\\
& + \frac{1}{M^2}\sum_{\substack{m,n=1\\m\neq n}}^{M}\left\lvert\hat{\mathbf{y}}_m\cdot\hat{\mathbf{y}}_n\right\rvert^2 S_{mn}(\mathbf{k}).
\end{aligned}
\label{eq:M_beam_spectrum_U_separate}
\end{equation}
The individual beam contributions ($m=n$) could be approximated using a Gaussian aperture, with its characteristic lengths chosen to reproduce the second order spatial statistics of the corresponding top-hat aperture, as given in Ref.~\cite[Table~II]{min2026statistical}, or calculated directly as
\begin{equation}
S_{mm}(\mathbf{k}) = \frac{16\pi k_0}{k_c^4k_{\perp m}}\sqrt{k_c^2-\left(\frac{k_{\perp m}}{2}+\frac{k_0|k_{\parallel m}|}{k_{\perp m}}\right)^2}.
\label{eq:spectrum_U_circular}
\end{equation}
for a circular top-hat aperture, where $k_c$ is defined in Eq.~\eqref{eq:wave_number_cutoff}, $k_{\perp m}^2=(\hat{\mathbf{x}}_m\cdot\mathbf{k})^2+(\hat{\mathbf{y}}_m\cdot\mathbf{k})^2$, and $k_{\parallel m}=\hat{\mathbf{z}}_m\cdot\mathbf{k}$. The spectrum vanishes wherever the square root argument is negative. For the interference contributions ($m\neq n$), the corresponding circular aperture spectrum is derived in Appendix~\ref{sec:Derivation_of_interference_spectra} for beams separated by the half angle $\gamma_{mn}$ satisfying $\sin(2\gamma_{mn})\gg 1/(2f_\#)$, giving
\begin{equation}
S_{mn}\approx\frac{8\pi\left[\max\{r_m+r_n,s\}-\max\{|r_m-r_n|,s\}\right]}{k_c^4\sin(2\gamma_{mn})},
\label{eq:M_beam_spectrum_interference}
\end{equation}
where $r_m$ and $r_n$ are defined as
\[
r_m := \sqrt{k_c^2-\frac{(\mathbf{q}\cdot\hat{\mathbf{z}}_n)^2}{\sin^2(2\gamma_{mn})}},\quad
r_n := \sqrt{k_c^2-\frac{(\mathbf{q}\cdot\hat{\mathbf{z}}_m)^2}{\sin^2(2\gamma_{mn})}},
\]
and $s:=|\mathbf{q}\cdot(\hat{\mathbf{z}}_m\times\hat{\mathbf{z}}_n)|/\sin(2\gamma_{mn})$ for $\mathbf{q} := \mathbf{k}-\kappa_0(\hat{\mathbf{z}}_m-\hat{\mathbf{z}}_n)$. The spectrum vanishes when either square-root argument is negative, while the maximum functions enforce the remaining spectral support. Each interference spectrum is centred at $\mathbf{k}=\kappa_0(\hat{\mathbf{z}}_m-\hat{\mathbf{z}}_n)$, where $\mathbf{q}=0$ and hence $r_m=r_n=k_c$ and $s=0$. The interference spectral density therefore attains its maximum
\begin{equation}
S_{mn}^{\max}=\frac{16\pi}{k_c^3\sin(2\gamma_{mn})}, \quad \text{for } m\neq n.
\end{equation}

These results establish the intensity spectra for the OMEGA configuration illustrated in Fig.~\ref{fig:M_crossing_beams} (b), which will be used when considering the electron density response. For square top-hat apertures, such as the NIF configurations illustrated in Fig.~\ref{fig:M_crossing_beams} (a), the anisotropic geometry complicates even the individual beam spectra, and the relative aperture orientations further complicate the interference spectra. We therefore approximate the spectra using circular-aperture expressions.

\subsection{\label{sec:Density_response}Density response}
Randomized laser fields can drive electron density fluctuations through stochastic ponderomotive force \cite{brantov1999plasma,grech2006plasma}. These fluctuations depend on both the spatial structure of the laser intensity and the plasma response. The spectral distribution is of particular interest, since electron density fluctuations can be probed through Thomson scattering~\cite{froula2010plasma}. The density response spectrum to a single speckle beam had been examined in Ref.~\cite{min2026statistical}, and having established the intensity spectrum for an arbitrary number of overlapping beams, we can now extend this analysis to the multi-beam case.

For electron density response $\delta n_e$, its spatial spectral density $S_{\delta n_e\delta n_e}\colon \mathbb{R}^3\to\mathbb{R}_{\geq0}$, can be written as
\begin{equation}
S_{\delta n_e\delta n_e}(\mathbf{k}) = \frac{n_{e0}^2\langle I \rangle^2}{4c^2T_e^2n_c^2} \left\lvert \mathcal{H}(\mathbf{k}) \right\rvert^2 S_{UU}(\mathbf{k}),
\label{eq:density_fluctuations}
\end{equation}
such that
\[
\left\langle \delta \hat{n}_e(\mathbf{k})\overline{\delta \hat{n}_e(\mathbf{k}^\prime)}\right\rangle = (2\pi)^3\, S_{\delta n_e\delta n_e}(\mathbf{k}) \delta(\mathbf{k}-\mathbf{k}^\prime),
\]
where $n_{e0}$ denotes the equilibrium electron density, $T_e$ denotes the electron temperature in energy units, $n_c=\varepsilon_0m_e\omega_0^2/q_e^2$ denotes the critical electron density for electron mass $m_e$ and charge $q_e$. The response function $\hat{\mathcal{H}}$ is defined as
\[
\left\lvert \mathcal{H}(\mathbf{k}) \right\rvert^2 := \left(\frac{k^2+\lambda_{Di}^{-2}}{k^2+\lambda_{De}^{-2}+\lambda_{Di}^{-2}}\right)^2
\]
in the kinetic description~\cite{min2026statistical} and $\left\lvert \mathcal{H}(\mathbf{k}) \right\rvert^2=1$ in the fluid description~\cite{brantov1999plasma}. Then substituting Eq.~\eqref{eq:M_beam_spectrum_U_separate} into Eq.~\eqref{eq:density_fluctuations} gives us the spatial spectral density for the electron density response due to $M$ overlapping speckled beams
\begin{equation}
\begin{aligned}
S_{\delta n_e\delta n_e}(\mathbf{k}) ={}& \frac{n_{e0}^2\langle I \rangle^2}{4c^2T_e^2n_c^2M^2} \left\lvert \mathcal{H}(\mathbf{k}) \right\rvert^2 \Bigg(\sum_{m=1}^{M}S_{mm}(\mathbf{k})\\
& + \sum_{\substack{m,n=1\\m\neq n}}^{M}\left\lvert\hat{\mathbf{y}}_m\cdot\hat{\mathbf{y}}_n\right\rvert^2 S_{mn}(\mathbf{k})\Bigg),
\end{aligned}
\label{eq:density_fluctuations_expand}
\end{equation}
where $S_{mm}$ and $S_{mn}$ are defined in Eqs.~\eqref{eq:spectrum_U_circular} and~\eqref{eq:M_beam_spectrum_interference} respectively.

Eq.~\eqref{eq:density_fluctuations_expand} separates the electron density fluctuation spectrum into contributions from individual beams and interference between pairs of crossing beams. The individual beam contributions are concentrated near the origin of $\mathbf{k}$ space, whereas interference between beams $m$ and $n$ introduces spectral features centred at $\mathbf{k}_{mn}=\kappa_0(\hat{\mathbf{z}}_m-\hat{\mathbf{z}}_n)$. The positions of the spectral features depend only on the beam propagation directions, while their polarization overlap $|\hat{\mathbf{y}}_m\cdot\hat{\mathbf{y}}_n|^2$ determine the amplitude.

Fig.~\ref{fig:Density_spectrum} shows the electron density fluctuation spectral densities calculated from Eq.~\eqref{eq:density_fluctuations_expand} for the NIF and OMEGA configurations illustrated in Fig.~\ref{fig:M_crossing_beams} (a) and (b), respectively, assuming all beams intersect at a single point. Panels (a,b) show the three-dimensional spectra with the region near the origin excluded to emphasize the contributions from interference between crossing beams. Panels (c,d) show the corresponding spectra in the $k_y=0$ plane, showing the low wavevector electron density fluctuations driven by speckles within individual beams, independently of interference between crossing beams. For the NIF configuration, the angular gap between beam propagation directions produces a corresponding gap in the spectral density, while the more uniformly distributed OMEGA beams produce a more uniform spectrum.

\section{\label{sec:Conclusion}Conclusion}
After formally defining speckles, establishing their effective sizes, and introducing the Kac-Rice formula for counting intensity maxima, we analyzed the statistical properties of speckles due to multiple crossing laser beams. For two statistically independent crossing beams, we derived the intensity probability density function, while Lemma~\ref{lem:speckle_size} and Theorem~\ref{the:speckle_abundance} give the effective speckle sizes and the asymptotic abundance of high-intensity speckles, respectively. The theoretical predictions for the asymptotic abundance of high intensity speckles were found to agree well with Monte Carlo simulations. The two speckled beam results were then applied to a symmetric quad configuration, where Corollary~\ref{cor:4_speckle_abundance} established the asymptotic abundance of high intensity speckles. We further showed that the mean SBS reflectivity associated with saturated speckles decreases with increasing crossing angle in the high-intensity asymptotic regime. The decrease arises from the shortening of the longitudinal speckle length, which raises the intensity threshold required for local SBS saturation. For an arbitrary number of crossing beams, we derived the one-point and two-point intensity statistics using a coordinate free formulation. These expressions were then used to obtain the spectral density of laser driven electron density fluctuations for the NIF and OMEGA beam configurations. The spectra exhibited additional contributions arising from interference between pairs of crossing beams, which were absent from the single beam spectrum.

\begin{acknowledgments}
We gratefully acknowledge K.R. McMillen for helpful discussions and providing parameters used in OMEGA. I.D. Min-Roberts and W. Rozmus would like to acknowledge support from the U.S. Department of Energy (National Nuclear Security Administration) under Award No. DE-NA0004144: University of Rochester “National Inertial Confinement Fusion Program”. Work performed under the auspices of the U.S. Department of Energy by the Lawrence Livermore National Laboratory (LLNL) under Contract No. DE-AC52-07NA27344.
\end{acknowledgments}

\appendix
\section{\label{sec:Probability_density_function}Probability density function of a weighted sum of $\chi_2^2$ variables}
We derive the probability density function of a random field $U\colon\mathbb{R}^N\to[0,\infty)$ with one-point distribution defined as a weighted sum of independent chi-squared random variables,
\begin{equation}
U=\frac{1}{2}\sum_{q=1}^{r}w_q\chi_{2,q}^{2},
\label{eq:intensity_distribution}
\end{equation}
where $r$ is the rank of the weighted sum corresponding to the number of nonzero weights $w_q>0$, and for any fixed spatial point the $\chi_{2,q}^{2}$ are mutually independent chi-squared random variables with two degrees of freedom. Defining
\[
X_q:=\frac{w_q}{2}\chi_{2,q}^{2}, \implies U=\sum_{q=1}^{r}X_q,
\]
and since $\chi_2^2/2$ is exponentially distributed with unit mean, each $X_q$ is exponentially distributed with mean $w_q$, with probability density
\begin{equation}
f_{X_q}(u)=\frac{1}{w_q}\exp\left(-\frac{u}{w_q}\right),\qquad u\geq0.
\label{eq:modal_intensity_pdf}
\end{equation}
$X_q$ for all $q\in[1,r]$ are mutually independent, thus the probability density of $U$ is given by their repeated convolution,
\begin{equation}
f_U(u)=\mathop{\ast}_{q=1}^{r}f_{X_q}(u)=f_{X_1}*\cdots*f_{X_r}.
\label{eq:convolution_pdf}
\end{equation}
Taking the Laplace transform of Eq.~\eqref{eq:convolution_pdf} converts the convolution into a product and after using Eq.~\eqref{eq:modal_intensity_pdf} we obtain
\begin{equation}
\mathcal{L}\{f_U\}(s)=\prod_{q=1}^{r}\frac{1}{1+s w_q}.
\label{eq:intensity_laplace}
\end{equation}
Then by partial fraction decomposition, the product can be written as
\[
\prod_{q=1}^{r}\frac{1}{1+sw_q}=\sum_{q=1}^{r}\left[\prod_{\substack{p=1\\p\neq q}}^{r}\frac{w_q}{w_q-w_p}\right]\frac{1}{1+sw_q},
\]
assuming distinct weights $w_1,\ldots,w_r$. Inverting the Laplace transform then gives
\begin{equation}
f_U(u)=\sum_{q=1}^{r}\left[\prod_{\substack{p=1\\p\neq q}}^{r}\frac{w_q}{w_q-w_p}\right]\frac{1}{w_q}\exp\left(-\frac{u}{w_q}\right),
\label{eq:intensity_pdf_distinct_weights}
\end{equation}
for $u\geq0$. If two or more weights are identical, the probability density can be obtained by taking the appropriate limit of Eq.~\eqref{eq:intensity_pdf_distinct_weights}.

\section{\label{sec:Proof_of_Lemma}Proof of Lemma \ref{lem:speckle_size}}
We introduce the coordinate system $\boldsymbol{\xi}=(\varrho,\vartheta,\varphi)$, where $\varrho\colon\mathbb{R}^3\to[0,\infty)$, $\vartheta\colon\mathbb{R}^3\to[0,\pi/2]$, and $\varphi\colon\mathbb{R}^3\to[0,2\pi)$, such that
\begin{equation}
\boldsymbol{\xi}:=\begin{pmatrix}\varrho\\ \vartheta\\ \varphi\end{pmatrix}=\begin{pmatrix}\frac{1}{\sigma^2}(|\mathring{\mathcal{A}}_1|^2+|\mathring{\mathcal{A}}_2|^2)\\[2mm] \tan^{-1}\!\left(|\mathring{\mathcal{A}}_1|/|\mathring{\mathcal{A}}_2|\right)\\[2mm] \arg\!\left(\overline{\mathring{\mathcal{A}}_1}\mathring{\mathcal{A}}_2e^{i\Delta\psi}\right)\end{pmatrix},
\label{eq:coordinate_changes}
\end{equation}
with the joint probability measure
\begin{equation}
d\nu(\boldsymbol{\xi}) = \frac{\varrho}{2\pi}e^{-\varrho}\sin(2\vartheta)\,d\varrho\,d\vartheta\,d\varphi.
\label{eq:internal_coordinate_measure}
\end{equation}
Then to calculate the second derivative matrix of Eq.~\eqref{eq:crossing_beam_intensity} we can invert the coordinate system so that $|\mathring{\mathcal{A}}_1|=\sigma\sqrt{\varrho}\sin\vartheta$, $|\mathring{\mathcal{A}}_2|=\sigma\sqrt{\varrho}\cos\vartheta$, and $\overline{\mathring{\mathcal{A}}_1}\mathring{\mathcal{A}}_2e^{i\Delta\psi}=\frac{\sigma^2\varrho}{2}\sin(2\vartheta)e^{i\varphi}$, which implies
\begin{equation}
U=\frac{\varrho}{2}\left[1+\alpha\sin(2\vartheta)\cos\varphi\right]=\varrho F(\vartheta,\varphi).
\label{eq:U_internal_coordinates}
\end{equation}
Then at a local maximum $\mathbf{x}_*$, the critical point condition $\nabla U=F\nabla\varrho+\varrho\nabla F=0$ gives $\nabla\varrho=-(\varrho/F)\nabla F$, so differentiating Eq.~\eqref{eq:U_internal_coordinates} twice and applying this condition yields
\begin{equation}
\nabla\nabla U=F\nabla\nabla \varrho+\varrho\nabla\nabla F-\frac{2\varrho}{F}\nabla F(\nabla F)^\top.
\label{eq:hessian_internal_coordinates}
\end{equation}

For $0<\alpha\leq1$, the Hessian in Eq.~\eqref{eq:hessian_internal_coordinates} can be expressed in the high-intensity asymptotic form
\begin{equation}
\nabla\nabla U=F_*\nabla\nabla \varrho+\varrho\nabla\nabla F_*+O_{\mathrm{p}}\!\left(\sqrt{\varrho}\right),
\label{eq:hessian_at_saddle_point}
\end{equation}
where $F_*=(1+\alpha)/2$ is the maximum of $F$, attained at $\vartheta_*=\pi/4$ and $\varphi_*=0$, for which $\nabla F_*=0$. The evaluation at these angular values follows from conditioning on $U=u$. Using Eq.~\eqref{eq:internal_coordinate_measure}, the conditional angular density is proportional to $\sin(2\vartheta)F^{-2}\exp[-u/F(\vartheta,\varphi)]$, which concentrates at the maximum of $F$ as $u\to\infty$. The angular fluctuations are $O_{\mathrm{p}}(u^{-1/2})$, and their contributions to the Hessian are absorbed into the remainder in Eq.~\eqref{eq:hessian_at_saddle_point}. Then by direct calculation, $\partial_{\vartheta\vartheta} F_* = -2\alpha$, $\partial_{\varphi\varphi} F_* = -\alpha/2$, and $\partial_{\vartheta\varphi}F_* = 0$, so that
\begin{equation}
\nabla\nabla F_*=-2\alpha\,\nabla\vartheta_*(\nabla\vartheta_*)^\top-\frac{\alpha}{2}\nabla\varphi_*(\nabla\varphi_*)^\top.
\label{eq:nabla2F_crit}
\end{equation}
To calculate the coordinate gradients, write the two vector potential envelopes as
\[
\mathring{\mathcal{A}}_1=\sigma\sqrt{\varrho}\sin\vartheta\,e^{i\varphi_1},\qquad \mathring{\mathcal{A}}_2=\sigma\sqrt{\varrho}\cos\vartheta\,e^{i\varphi_2},
\]
such that $\varphi=\varphi_2-\varphi_1+\Delta\psi$, and define
\[
\mathbf{X}_j:=\Re\!\left(e^{-i\varphi_j}\nabla\mathring{\mathcal{A}}_j\right),\qquad \mathbf{Y}_j:=\Im\!\left(e^{-i\varphi_j}\nabla\mathring{\mathcal{A}}_j\right).
\]
Then the gradients can be expressed as
\begin{align}
\nabla\varrho &= \frac{2\sqrt{\varrho}}{\sigma}\left(\sin\vartheta\,\mathbf{X}_1+\cos\vartheta\,\mathbf{X}_2\right),
\label{eq:grad_rho}\\
\nabla\vartheta &= \frac{1}{\sigma\sqrt{\varrho}}\left(\cos\vartheta\,\mathbf{X}_1-\sin\vartheta\,\mathbf{X}_2\right),
\label{eq:grad_theta}\\
\nabla\varphi &= \nabla\Delta\psi+\frac{1}{\sigma\sqrt{\varrho}}\left(\frac{\mathbf{Y}_2}{\cos\vartheta}-\frac{\mathbf{Y}_1}{\sin\vartheta}\right).
\label{eq:grad_phi}
\end{align}
Since $\mathbf{X}_j$ and $\mathbf{Y}_j$ remain $O_{\mathrm{p}}(1)$ under high-amplitude conditioning, it follows that
\[
\nabla\vartheta_*=O_{\mathrm{p}}(\varrho^{-1/2}),\qquad \nabla\varphi_*=\nabla\Delta\psi+O_{\mathrm{p}}(\varrho^{-1/2}).
\]
From Eq.~\eqref{eq:crossing_beam_intensity}, $\Delta\psi=-2\kappa_0x\sin\gamma$ and $(\nabla\Delta\psi)(\nabla\Delta\psi)^\top=4\kappa_0^2\sin^2\gamma\,\hat{\mathbf{x}}\otimes\hat{\mathbf{x}}$. Substituting into Eq.~\eqref{eq:nabla2F_crit} gives
\begin{equation}
\varrho\nabla\nabla F_*=-2\alpha\varrho\kappa_0^2\sin^2\gamma\,\hat{\mathbf{x}}\otimes\hat{\mathbf{x}}+O_{\mathrm{p}}\!\left(\sqrt{\varrho}\right).
\label{eq:F_hessian_asymptotic}
\end{equation}
Differentiating Eq.~\eqref{eq:grad_rho} and using $\langle\nabla\nabla \mathring{\mathcal{A}}_j\mid\mathring{\mathcal{A}}_j\rangle = -\mathring{\mathbf{m}}_j^{(2)}\mathring{\mathcal{A}}_j$, with the gradient product terms of lower order, gives
\begin{equation}
F_*\nabla\nabla \varrho =-\frac{(1+\alpha)\varrho}{2} \left(\mathring{\mathbf{m}}_1^{(2)} + \mathring{\mathbf{m}}_2^{(2)}\right) + O_{\mathrm{p}}\!\left(\sqrt{\varrho}\right),
\label{eq:rho_hessian_at_saddle}
\end{equation}
at $\vartheta_*=\pi/4$. Substituting Eqs.~\eqref{eq:F_hessian_asymptotic} and~\eqref{eq:rho_hessian_at_saddle} in Eq.~\eqref{eq:hessian_at_saddle_point} yields
\begin{equation}
\begin{aligned}
\nabla\nabla U(\mathbf{x}_*)={}&-\frac{\varrho}{2}\left[(1+\alpha)\left(\mathring{\mathbf{m}}_1^{(2)}+\mathring{\mathbf{m}}_2^{(2)}\right)\right.\\
&\left.+4\alpha\kappa_0^2\sin^2\gamma\,\hat{\mathbf{x}}\otimes\hat{\mathbf{x}}\right]+O_{\mathrm{p}}\!\left(\sqrt{\varrho}\right).
\end{aligned}
\label{eq:hessian_U}
\end{equation}
Since $U(\mathbf{x}_*)=\varrho F_*=(1+\alpha)\varrho/2$ to leading order, we obtain
\begin{equation}
\begin{aligned}
-\frac{\nabla\nabla U(\mathbf{x}_*)}{U(\mathbf{x}_*)}={}&\mathring{\mathbf{m}}_1^{(2)} + \mathring{\mathbf{m}}_2^{(2)} +\frac{4\alpha\kappa_0^2}{1+\alpha}\sin^2\gamma\\
& \times\hat{\mathbf{x}}\otimes\hat{\mathbf{x}} + O_{\mathrm{p}}\!\left(U(\mathbf{x}_*)^{-1/2}\right).
\end{aligned}
\label{eq:hessian_U_normalized}
\end{equation}
Comparison with Eq.~\eqref{eq:effective_second_moment} establishes the effective second-order spectral moment in Eq.~\eqref{eq:2_effective_second_order_spectral_moment}.

For $\alpha=0$, $F=1/2$ is independent of the angular coordinates, and Eq.~\eqref{eq:hessian_internal_coordinates} reduces to
\[
\nabla\nabla U = -\varrho\left(\sin^2\vartheta\,\mathring{\mathbf{m}}_1^{(2)}+\cos^2\vartheta\,\mathring{\mathbf{m}}_2^{(2)}\right)+O_{\mathrm{p}}\!\left(\sqrt{\varrho}\right).
\]
Since $2U=\varrho$, normalization gives
\begin{equation}
\begin{aligned}
-\frac{\nabla\nabla U(\mathbf{x}_*)}{2U(\mathbf{x}_*)}={}&\sin^2\vartheta\,\mathring{\mathbf{m}}_1^{(2)}+\cos^2\vartheta\,\mathring{\mathbf{m}}_2^{(2)}\\
&+O_{\mathrm{p}}(U(\mathbf{x}_*)^{-\frac{1}{2}}),
\end{aligned}
\end{equation}
and comparison with Eq.~\eqref{eq:effective_second_moment} gives Eq.~\eqref{eq:two_beam_effective_moment_alpha0} which completes the proof.

\section{\label{sec:Proof_of_Theorem}Proof of Theorem \ref{the:speckle_abundance}}
Using the coordinate system $\boldsymbol{\xi}=(\varrho,\vartheta,\varphi)$ defined in Eq.~\eqref{eq:coordinate_changes}, with $U=\varrho F(\vartheta,\varphi)$ as in Eq.~\eqref{eq:U_internal_coordinates}, the Kac-Rice formula from Eq.~\eqref{eq:kac_rice} can be expressed as
\begin{equation}
\begin{aligned}
\left\langle M_\Omega^+(u)\right\rangle =& |\Omega|\int \left\langle \left|\det(\nabla^2U)\right| \mathbf{1}_{\{\nabla^2U\prec0\}} \,\middle|\, \nabla U=0,\boldsymbol{\xi} \right\rangle\\
&\times \mathbf{1}_{\{U(\boldsymbol{\xi})>u\}}\, f_{\nabla U\mid\boldsymbol{\xi}}(0)\,d\nu(\boldsymbol{\xi}),
\end{aligned}
\label{eq:metatheorem_IAP_coordinates}
\end{equation}
where the induced probability measure is defined in Eq.~\eqref{eq:internal_coordinate_measure}. We therefore need the conditional density of $\nabla U$ and the conditional expectation involving the Hessian determinant.

To calculate $f_{\nabla U\mid\boldsymbol{\xi}}$, we note that conditioning on $\boldsymbol{\xi}$ fixes the magnitudes and relative phase of the two beam envelopes, leaving only a common phase. By circular Gaussian symmetry, this common phase does not affect the distribution of $\nabla U$, which is an affine function of the Gaussian envelope derivatives. Hence $\nabla U\mid\boldsymbol{\xi}\sim\mathcal{N}(\boldsymbol{\mu}(\boldsymbol{\xi}),\mathbf{C}(\boldsymbol{\xi}))$, with
\begin{equation}
f_{\nabla U\mid\boldsymbol{\xi}}(0)=\frac{\exp\!\left[-\frac{1}{2}\boldsymbol{\mu}^{\top}\mathbf{C}^{-1}\boldsymbol{\mu}\right]}{(2\pi)^{N/2}\sqrt{\det\mathbf{C}}},
\label{eq:conditional_gradient_density}
\end{equation}
where $\boldsymbol{\mu}(\boldsymbol{\xi}):=\langle\nabla U\mid\boldsymbol{\xi}\rangle$ and $\mathbf{C}(\boldsymbol{\xi}):=\operatorname{Cov}(\nabla U\mid\boldsymbol{\xi})$. To calculate the mean, we differentiate Eq.~\eqref{eq:U_internal_coordinates}
\begin{equation}
\nabla U=F\nabla\varrho+\varrho F_\vartheta\nabla\vartheta+\varrho F_\varphi\nabla\varphi.
\label{eq:grad_U_chain_rule}
\end{equation}
The first-order spectral moments of the phase-corrected envelopes vanish meaning $\langle\nabla\varrho\mid\boldsymbol{\xi}\rangle=\langle\nabla\vartheta\mid\boldsymbol{\xi}\rangle=0$ and $\langle\nabla\varphi\mid\boldsymbol{\xi}\rangle=\nabla\Delta\psi$, therefore,
\begin{equation}
\boldsymbol{\mu}(\boldsymbol{\xi})=\varrho F_\varphi\nabla\Delta\psi=-\frac{\alpha\varrho}{2}\sin(2\vartheta)\sin\varphi\,\nabla\Delta\psi.
\label{eq:mean_grad}
\end{equation}
For the covariance, we substitute Eqs.~\eqref{eq:grad_rho}, \eqref{eq:grad_theta}, \eqref{eq:grad_phi} into Eq.~\eqref{eq:grad_U_chain_rule} and subtracting the conditional mean
\begin{align}
\nabla U-\boldsymbol{\mu}=\frac{\sqrt{\varrho}}{\sigma}\Bigg[&\left(2F\sin\vartheta+F_\vartheta\cos\vartheta\right)\mathbf{X}_1\nonumber\\
&+\left(2F\cos\vartheta-F_\vartheta\sin\vartheta\right)\mathbf{X}_2\\
&-\frac{F_\varphi}{\sin\vartheta}\mathbf{Y}_1+\frac{F_\varphi}{\cos\vartheta}\mathbf{Y}_2\Bigg]\nonumber.
\end{align}
For independent proper complex Gaussian phase-corrected envelopes,
\begin{equation}
\left\langle\mathbf{X}_j\mathbf{X}_j^{\top}\right\rangle=\left\langle\mathbf{Y}_j\mathbf{Y}_j^{\top}\right\rangle=\frac{\sigma^2}{2}\mathring{\mathbf{m}}_j^{(2)},\quad \left\langle\mathbf{X}_j\mathbf{Y}_j^{\top}\right\rangle=0.
\end{equation}
Thus
\begin{equation}
\begin{aligned}
\mathbf{C}(\boldsymbol{\xi}) &={} \frac{\varrho}{2}\left[\left(2F\sin\vartheta+F_\vartheta\cos\vartheta\right)^2+\frac{F_\varphi^2}{\sin^2\vartheta}\right]\mathring{\mathbf{m}}_1^{(2)} \\
& + \frac{\varrho}{2}\left[\left(2F\cos\vartheta-F_\vartheta\sin\vartheta\right)^2+\frac{F_\varphi^2}{\cos^2\vartheta}\right]\mathring{\mathbf{m}}_2^{(2)}.
\end{aligned}
\label{eq:cov_grad_expanded}
\end{equation}
Since both the conditional mean and covariance are linear in $\varrho$, we can write
\begin{equation}
\boldsymbol{\mu}(\boldsymbol{\xi})=\varrho\boldsymbol{\mu}_0(\vartheta,\varphi),\qquad \mathbf{C}(\boldsymbol{\xi})=\varrho\mathbf{C}_0(\vartheta,\varphi),
\label{eq:mean_and_cov_varrho}
\end{equation}
which reduces Eq.~\eqref{eq:conditional_gradient_density} to
\begin{equation}
f_{\nabla U\mid\boldsymbol{\xi}}(0)=\frac{\exp\!\left[-\frac{\varrho}{2}\boldsymbol{\mu}_0^{\top}\mathbf{C}_0^{-1}\boldsymbol{\mu}_0\right]}{(2\pi\varrho)^{N/2}\sqrt{\det\mathbf{C}_0}}.
\label{eq:conditional_gradient_density_scaled}
\end{equation}

Now we calculate the conditional Hessian determinant in Eq.~\eqref{eq:metatheorem_IAP_coordinates}. Under Gaussian conditioning on $\boldsymbol{\xi}$ and $\nabla U=0$, the Hessian has the asymptotic form
\begin{equation}
\nabla\nabla U=\varrho\mathbf{H}(\vartheta,\varphi)+O_{\mathrm{p}}\!\left(\sqrt{\varrho}\right),
\label{eq:hessian_leading_order}
\end{equation}
where $\mathbf{H}(\vartheta,\varphi)$ is the deterministic leading order matrix. Provided $\mathbf{H}$ is nonsingular,
\begin{equation}
\det(\nabla\nabla U)=\varrho^N\det\mathbf{H}\left[1+O_{\mathrm{p}}\!\left(\varrho^{-1/2}\right)\right].
\label{eq:det_hessian_leading_order}
\end{equation}
The negative-definiteness condition is therefore determined by $\mathbf{H}$ to leading order.

Thus substituting Eqs.~\eqref{eq:conditional_gradient_density_scaled} and \eqref{eq:det_hessian_leading_order} into Eq.~\eqref{eq:metatheorem_IAP_coordinates} gives
\begin{widetext}
\begin{equation}
\begin{aligned}
\left\langle M_\Omega^+(u)\right\rangle={}&\frac{|\Omega|}{(2\pi)^{\frac{N}{2}+1}}\int_0^{\pi/2}\int_0^{2\pi}\frac{|\det\mathbf{H}(\vartheta,\varphi)|\,\mathbf{1}_{\{\mathbf{H}\prec0\}}\sin(2\vartheta)}{\sqrt{\det\mathbf{C}_0(\vartheta,\varphi)}}\\
&\times\int_{u/F(\vartheta,\varphi)}^\infty\varrho^{\frac{N}{2}+1}\exp\!\left[-\varrho\left(1+\frac{1}{2}\boldsymbol{\mu}_0^{\top}\mathbf{C}_0^{-1}\boldsymbol{\mu}_0\right)\right]\left[1+O\!\left(\varrho^{-1/2}\right)\right]\,d\varrho\,d\varphi\,d\vartheta.
\end{aligned}
\label{eq:metatheorem_IAP_pre_asymp}
\end{equation}
\end{widetext}
The $\varrho$ integral can then be evaluated asymptotically at its lower endpoint, giving
\begin{equation}
\begin{aligned}
\left\langle M_\Omega^+(u)\right\rangle={}&\frac{|\Omega|u^{N/2+1}}{(2\pi)^{N/2+1}}\int_0^{\pi/2}\int_0^{2\pi}W(\vartheta,\varphi)\\
&\times e^{-u\Psi(\vartheta,\varphi)}\left[1+O\!\left(u^{-1/2}\right)\right]\,d\varphi\,d\vartheta,
\end{aligned}
\label{eq:pre_steepest_descent}
\end{equation}
where
\begin{equation}
\Psi(\vartheta,\varphi)=\frac{1}{F(\vartheta,\varphi)}\left(1+\frac{1}{2}\boldsymbol{\mu}_0^{\top}\mathbf{C}_0^{-1}\boldsymbol{\mu}_0\right)
\label{eq:phase_function}
\end{equation}
and
\begin{equation}
W(\vartheta,\varphi)=\frac{|\det\mathbf{H}|\,\mathbf{1}_{\{\mathbf{H}\prec0\}}\sin(2\vartheta)}{\sqrt{\det\mathbf{C}_0}\,F^{\frac{N}{2}+1}\left(1+\frac{1}{2}\boldsymbol{\mu}_0^{\top}\mathbf{C}_0^{-1}\boldsymbol{\mu}_0\right)}.
\label{eq:amplitude_function}
\end{equation}
Steepest descent method tells us that, in the high threshold limit $u\to\infty$, the dominant contribution to the angular integral comes from the point $(\vartheta_*,\varphi_*)$ at which
\[
\Psi(\vartheta_*,\varphi_*) = \min_{\vartheta,\varphi}\Psi(\vartheta,\varphi).
\]
For $0<\alpha\leq1$, we have $\boldsymbol{\mu}_0^{\top}\mathbf{C}_0^{-1}\boldsymbol{\mu}_0\geq0$, while $F$ is maximized at $\vartheta_*=\pi/4$ and $\varphi_*=0\pmod{2\pi}$, where $\boldsymbol{\mu}_0=0$. Therefore,
\begin{equation}
\Psi_*=\Psi(\vartheta_*,\varphi_*)=\frac{2}{1+\alpha}.
\end{equation}
For $\alpha=0$, however, $F=1/2$ and $\boldsymbol{\mu}_0=\mathbf{0}$ for all angular coordinates, so $\Psi=2$ throughout the domain. The minimum is no longer isolated, and the two cases must be treated separately.\\

\textbf{Case 1:} $0<\alpha\leq1$.

For fixed $0<\alpha\leq1$, $\Psi$ has a nondegenerate minimum at $(\vartheta_*,\varphi_*)$, so applying steepest descent method to Eq.~\eqref{eq:pre_steepest_descent} gives
\begin{equation}
\begin{aligned}
\left\langle M_\Omega^+(u)\right\rangle={}&\frac{|\Omega|u^{N/2}}{(2\pi)^{N/2}}\frac{W(\vartheta_*,\varphi_*)}{\sqrt{\det\nabla\nabla_{\vartheta,\varphi}\Psi_*}}\\
&\times e^{-u\Psi_*}\left[1+O\!\left(u^{-1/2}\right)\right].
\end{aligned}
\label{eq:maxima_steepest_descent}
\end{equation}
Direct differentiation of Eq.~\eqref{eq:phase_function} and evaluation of Eq.~\eqref{eq:cov_grad_expanded} at $(\vartheta_*,\varphi_*)$ gives
\begin{equation}
\begin{aligned}
\det\nabla\nabla_{\vartheta,\varphi}\Psi_*={}&\frac{16\alpha^2}{(1+\alpha)^4}\Bigg[1+\frac{\alpha}{1+\alpha}(\nabla\Delta\psi)^{\top}\\
&\times\left(\mathring{\mathbf{m}}_1^{(2)}+\mathring{\mathbf{m}}_2^{(2)}\right)^{-1}(\nabla\Delta\psi)\Bigg].
\end{aligned}
\label{eq:phase_hessian_determinant}
\end{equation}
Eq.~\eqref{eq:amplitude_function} at point $(\vartheta_*,\varphi_*)$ is
\begin{equation}
W(\vartheta_*,\varphi_*)=\frac{2^{\frac{3}{2}N+1}\left|\det\mathbf{H}_*\right|\,\mathbf{1}_{\{\mathbf{H}_*\prec0\}}}{(1+\alpha)^{\frac{3}{2}N+1}\sqrt{\det\!\left(\mathring{\mathbf{m}}_1^{(2)}+\mathring{\mathbf{m}}_2^{(2)}\right)}},
\label{eq:g_saddle}
\end{equation}
where by Lemma~\ref{lem:speckle_size} $\mathbf{H}_*=-(1+\alpha)\boldsymbol{\Lambda}$. Then by the determinant identity $\det(\mathbf{A}+\mathbf{u}\mathbf{v}^{\top})=\det(\mathbf{A})\left(1+\mathbf{v}^{\top}\mathbf{A}^{-1}\mathbf{u}\right)$, we get
\begin{equation}
\begin{aligned}
\det(2\boldsymbol{\Lambda}) &= \det\left(\mathring{\mathbf{m}}_1^{(2)}+\mathring{\mathbf{m}}_2^{(2)}\right)\Bigg[1+\frac{\alpha}{1+\alpha}\\
&\times(\nabla\Delta\psi)^{\top}\left(\mathring{\mathbf{m}}_1^{(2)}+\mathring{\mathbf{m}}_2^{(2)}\right)^{-1}(\nabla\Delta\psi)\Bigg].
\end{aligned}
\end{equation}
Since $\boldsymbol{\Lambda}\succ0$, we have $\mathbf{H}_*\prec0$, so the negative-definiteness indicator equals one at the saddle point, therefore,
\begin{equation}
\frac{W(\vartheta_*,\varphi_*)}{\sqrt{\det\nabla\nabla_{\vartheta,\varphi}\Psi(\vartheta_*,\varphi_*)}}=\frac{2^{N-1}\sqrt{\det\boldsymbol{\Lambda}}}{\alpha(1+\alpha)^{N/2-1}}.
\end{equation}
Then substituting into Eq.~\eqref{eq:maxima_steepest_descent} gives Eq.~\eqref{eq:maxima_0alpha1}.\\

\textbf{Case 2:} $\alpha=0$.

For orthogonal polarizations, $F=1/2$ and $\boldsymbol{\mu}_0=\mathbf{0}$, giving $\Psi=2$ independently of $\vartheta$ and $\varphi$. The angular integral therefore does not localize about an isolated saddle point. Instead, we use the $\vartheta$-dependent effective second-order spectral moment obtained in Lemma~\ref{lem:speckle_size}, Eq.~\eqref{eq:two_beam_effective_moment_alpha0},
\begin{equation}
\boldsymbol{\Lambda}(\vartheta)=\sin^2\vartheta\,\mathring{\mathbf{m}}_1^{(2)}+\cos^2\vartheta\,\mathring{\mathbf{m}}_2^{(2)}.
\end{equation}
From Eq.~\eqref{eq:cov_grad_expanded}, $\mathbf{C}_0(\vartheta)=\boldsymbol{\Lambda}(\vartheta)/2$, while the lemma gives $\mathbf{H}(\vartheta)=-\boldsymbol{\Lambda}(\vartheta)$. Since $\boldsymbol{\Lambda}(\vartheta)\succ0$, Eq.~\eqref{eq:amplitude_function} reduces to
\begin{equation}
W(\vartheta)=2^{N+1}\sin(2\vartheta)\sqrt{\det\boldsymbol{\Lambda}(\vartheta)}.
\label{eq:amplitude_alpha0}
\end{equation}
Substituting into Eq.~\eqref{eq:pre_steepest_descent} and evaluating the $\varphi$ integral gives Eq.~\eqref{eq:maxima_alpha0}.\\

\textbf{Case 3:} $u\alpha=O(1)$.

The fixed-$\alpha$ asymptotic result is nonuniform as $\alpha\to0$, whereas the orthogonally polarized case has a different algebraic dependence on $u$. To connect the two limits, we consider the transition regime $\alpha\to0$ and $u\to\infty$ with $u\alpha=O(1)$. Expanding Eq.~\eqref{eq:phase_function} for small $\alpha$, using $\boldsymbol{\mu}_0=O(\alpha)$, gives
\begin{equation}
\Psi(\vartheta,\varphi)=2-2\alpha\sin(2\vartheta)\cos\varphi+O(\alpha^2).
\end{equation}
Since $\alpha=O(u^{-1})$ in this regime,
\begin{equation}
e^{-u\Psi(\vartheta,\varphi)}=e^{-2u}e^{2\alpha u\sin(2\vartheta)\cos\varphi}\left[1+O(u^{-1})\right].
\end{equation}
Similarly, the amplitude function satisfies $W(\vartheta,\varphi)=W_0(\vartheta)+O(u^{-1})$, where $W_0$ is given by Eq.~\eqref{eq:amplitude_alpha0}. The remaining phase integral is
\begin{equation}
\int_0^{2\pi}e^{2\alpha u\sin(2\vartheta)\cos\varphi}\,d\varphi=2\pi\mathcal{I}_0\!\left(2\alpha u\sin(2\vartheta)\right),
\end{equation}
where $\mathcal{I}_0$ is the modified Bessel function of the first kind of order zero. Substituting into Eq.~\eqref{eq:pre_steepest_descent} gives Eq.~\eqref{eq:maxima_transition} which completes the proof.

\section{\label{sec:Derivation_of_interference_spectra}Derivation of interference spectra}
We derive the interference spectral density $S_{mn}(\mathbf{k})$ for distinct beams ($m\neq n$) with circular top-hat apertures. Applying the Fourier shift theorem to Eq.~\eqref{eq:M_beam_spectrum_pair} gives
\begin{equation}
S_{mn}(\mathbf{k})=\frac{1}{\sigma^4}\mathcal{F}\left[C_{\mathring{\mathcal{A}}_m\mathring{\mathcal{A}}_m}\overline{C_{\mathring{\mathcal{A}}_n\mathring{\mathcal{A}}_n}}\right]\left(\mathbf{q}\right),
\label{eq:M_beam_spectrum_fourier_shift}
\end{equation}
where $\mathbf{q} := \mathbf{k}-\kappa_0(\hat{\mathbf{z}}_m-\hat{\mathbf{z}}_n)$. The spatial covariance for the circular aperture derived in Ref.~\cite[Eq. (34)]{min2026statistical} can be expressed in Cartesian coordinates as
\begin{equation}
\begin{aligned}
C_{\mathring{\mathcal A}_m\mathring{\mathcal A}_m}(\mathbf{x}) ={}& \frac{\sigma^2}{\pi k_c^2}\int_{|\mathbf{k}^{\prime}|\leq k_c} \exp\Bigg[i\mathbf{k}^{\prime}\cdot\mathbf{x}\\
& - \frac{i}{2k_0}\left(|\mathbf{k}^{\prime}|^2-\frac{k_c^2}{2}\right)z_m\Bigg]d^2\mathbf{k}^{\prime},
\end{aligned}
\label{eq:M_beam_covariance_circular}
\end{equation}
where $\mathbf{k}^{\prime}=k_{x,m}^{\prime}\hat{\mathbf{x}}_m+k_{y,m}^{\prime}\hat{\mathbf{y}}_m$. Substituting Eq.~\eqref{eq:M_beam_covariance_circular} into Eq.~\eqref{eq:M_beam_spectrum_fourier_shift} and evaluating the Fourier transform gives
\begin{equation}
\begin{aligned}
S_{mn}(\mathbf{k})={}&\frac{8\pi}{k_c^4}
\int_{|\mathbf{k}^{\prime}|\leq k_c}\int_{|\mathbf{k}^{\prime\prime}|\leq k_c} \delta^{(3)}\Bigg(\mathbf{q} -\mathbf{k}^{\prime} + \mathbf{k}^{\prime\prime}\\
&  + \frac{|\mathbf{k}^{\prime}|^2-k_c^2/2}{2k_0}\hat{\mathbf{z}}_m - \frac{|\mathbf{k}^{\prime\prime}|^2-k_c^2/2}{2k_0}\hat{\mathbf{z}}_n \Bigg)\\
& \times d^2\mathbf{k}^{\prime}\,d^2\mathbf{k}^{\prime\prime},
\end{aligned}
\label{eq:M_beam_spectrum_delta}
\end{equation}
where $\delta^{(3)}(\cdot)$ denotes the three dimensional Dirac delta function. For beams separated by an angle $2\gamma_{mn}$, the crossing geometry introduces spectral variations of order $k_c\sin(2\gamma_{mn})$, whereas the quadratic terms contribute corrections of order $k_c^2/k_0$. Thus, for $\sin(2\gamma_{mn})\gg k_c/k_0\simeq 1/(2f_\#)$, where $f_\#$ is the $f$-number, we neglect the quadratic terms to obtain
\begin{equation}
S_{mn}\approx\frac{8\pi}{k_c^4}\int_{|\mathbf{k}^{\prime}|\leq k_c}\int_{|\mathbf{k}^{\prime\prime}|\leq k_c}\!\!\delta(\mathbf{q}-\mathbf{k}^{\prime}+\mathbf{k}^{\prime\prime})\,d^2\mathbf{k}^{\prime}\,d^2\mathbf{k}^{\prime\prime}.
\end{equation}
Since $\mathbf{k}^{\prime\prime}$ lies in the plane perpendicular to $\hat{\mathbf{z}}_n$, integration over its two transverse components eliminates two of the three Dirac delta constraints. The remaining constraint requires $(\mathbf{k}^{\prime}-\mathbf{q})\cdot\hat{\mathbf{z}}_n=0$, giving
\begin{equation}
S_{mn}(\mathbf{k})\approx\frac{8\pi}{k_c^4}
\int_{\substack{|\mathbf{k}^{\prime}|\leq k_c\\|\mathbf{k}^{\prime}-\mathbf{q}|\leq k_c}}
\delta\left(\mathbf{k}^{\prime}\cdot\hat{\mathbf{z}}_n-\mathbf{q}\cdot\hat{\mathbf{z}}_n\right)d^2\mathbf{k}^{\prime}.
\end{equation}
The aperture bounds $|\mathbf{k}^{\prime}|\leq k_c$ and $|\mathbf{k}^{\prime}-\mathbf{q}|\leq k_c$ restrict the remaining integration perpendicular to the crossing plane, with corresponding half widths
\[
r_m=\sqrt{k_c^2-\frac{(\mathbf{q}\cdot\hat{\mathbf{z}}_n)^2}{\sin^2(2\gamma_{mn})}},\quad
r_n=\sqrt{k_c^2-\frac{(\mathbf{q}\cdot\hat{\mathbf{z}}_m)^2}{\sin^2(2\gamma_{mn})}},
\]
separated by $s=|\mathbf{q}\cdot(\hat{\mathbf{z}}_m\times\hat{\mathbf{z}}_n)|/\sin(2\gamma_{mn})$. Introducing the integration variable $s^\prime=\mathbf{k}^{\prime}\cdot(\hat{\mathbf{z}}_m\times\hat{\mathbf{z}}_n)/\sin(2\gamma_{mn})$ with the sign chosen so that the separation is positive, the aperture constraints become $|s^{\prime}|\leq r_m$, and $|s^{\prime}-s|\leq r_n$. Integrating the remaining Dirac delta function introduces a factor $1/\sin(2\gamma_{mn})$, giving
\begin{equation}
S_{mn}(\mathbf{k})\approx\frac{8\pi}{k_c^4\sin(2\gamma_{mn})} \int_{\max(-r_m,s-r_n)}^{\min(r_m,s+r_n)}ds^\prime,
\end{equation}
provided the upper limit exceeds the lower limit, and zero otherwise. Evaluating this integral yields Eq.~\eqref{eq:M_beam_spectrum_interference}.

\nocite{taylor2006gaussian,giacone1995angular}
\bibliography{references}

@book{michel2023introduction,
  title={Introduction to laser-plasma interactions},
  author={Michel, Pierre},
  year={2023},
  publisher={Springer Nature}
}

@article{garnier1999statistics,
  title={Statistics of the hot spots of smoothed beams produced by random phase plates revisited},
  author={Garnier, J},
  journal={Physics of plasmas},
  volume={6},
  number={5},
  pages={1601--1610},
  year={1999},
  publisher={American Institute of Physics}
}

@article{rose1993statistical,
  title={Statistical properties of laser hot spots produced by a random phase plate},
  author={Rose, Harvey A and DuBois, DF},
  journal={Physics of Fluids B: Plasma Physics},
  volume={5},
  number={2},
  pages={590--596},
  year={1993},
  publisher={American Institute of Physics}
}

@article{worsley1994local,
  title={Local maxima and the expected Euler characteristic of excursion sets of $\chi$ 2, F and t fields},
  author={Worsley, Keith J},
  journal={Advances in Applied Probability},
  volume={26},
  number={1},
  pages={13--42},
  year={1994},
  publisher={Cambridge University Press}
}

@book{adler2007random,
  title={Random fields and geometry},
  author={Adler, Robert J and Taylor, Jonathan E},
  year={2007},
  publisher={Springer}
}

@article{brantov1999plasma,
  title={Plasma fluctuations driven by a randomized laser beam},
  author={Brantov, AV and Bychenkov, V Yu and Tikhonchuk, VT and Rozmus, W and Senecha, VK},
  journal={Physics of Plasmas},
  volume={6},
  number={8},
  pages={3002--3011},
  year={1999},
  publisher={American Institute of Physics}
}

@article{rose1994laser,
  title={Laser hot spots and the breakdown of linear instability theory with application to stimulated Brillouin scattering},
  author={Rose, Harvey A and DuBois, DF},
  journal={Physical review letters},
  volume={72},
  number={18},
  pages={2883},
  year={1994},
  publisher={APS}
}

@article{kato1984random,
  title={Random phasing of high-power lasers for uniform target acceleration and plasma-instability suppression},
  author={Kato, Yoshiaki and Mima, K and Miyanaga, N and Arinaga, S and Kitagawa, Y and Nakatsuka, M and Yamanaka, C},
  journal={Physical Review Letters},
  volume={53},
  number={11},
  pages={1057},
  year={1984},
  publisher={APS}
}

@article{lehmberg1987theory,
  title={Theory of induced spatial incoherence},
  author={Lehmberg, RH and Schmitt, AJ and Bodner, SE},
  journal={Journal of applied physics},
  volume={62},
  number={7},
  pages={2680--2701},
  year={1987},
  publisher={American Institute of Physics}
}

@article{grech2006plasma,
  title={Plasma induced laser beam smoothing below the filamentation threshold},
  author={Grech, M and Tikhonchuk, VT and Riazuelo, G and Weber, S},
  journal={Physics of plasmas},
  volume={13},
  number={9},
  year={2006},
  publisher={AIP Publishing}
}

@book{froula2010plasma,
  title={Plasma Scattering of Electromagnetic Radiation: Theory and Measurement Techniques},
  author={Froula, Dustin H and Glenzer, Siegfried H and Luhmann Jr, Neville C and Sheffield, John},
  year={2010},
  publisher={Elsevier}
}

@article{min2026statistical,
  title = {Statistical analysis of speckle fields},
  author = {Min-Roberts, Ian D. and Rozmus, Wojciech and Michel, Pierre A.},
  journal = {Phys. Rev. E},
  volume = {114},
  issue = {4},
  pages = {045202},
  numpages = {13},
  year = {2026},
  month = {Oct},
  publisher = {American Physical Society},
  doi = {10.1103/dbsb-8lpt},
  url = {https://link.aps.org/doi/10.1103/dbsb-8lpt}
}

@article{borcea2018laser,
  title={Laser beam imaging from the speckle pattern of the off-axis scattered intensity},
  author={Borcea, Liliana and Garnier, Josselin},
  journal={SIAM Journal on Applied Mathematics},
  volume={78},
  number={2},
  pages={677--704},
  year={2018},
  publisher={SIAM}
}

@article{su20072d,
  title={2D light scattering patterns of mitochondria in single cells},
  author={Su, Xuan-Tao and Capjack, Clarence and Rozmus, Wojciech and Backhouse, Christopher},
  journal={Optics Express},
  volume={15},
  number={17},
  pages={10562--10575},
  year={2007},
  publisher={Optical Society of America}
}

@article{le2011statistical,
  title={Statistical spatial properties of speckle patterns generated by multiple laser beams},
  author={Le Cain, A and Riazuelo, G and Sajer, JM},
  journal={Physics of Plasmas},
  volume={18},
  number={8},
  year={2011},
  publisher={AIP Publishing}
}

@article{tsubakimoto1992suppression,
  title={Suppression of interference speckles produced by a random phase plate, using a polarization control plate},
  author={Tsubakimoto, K and Nakatsuka, M and Nakano, H and Kanabe, T and Jitsuno, T and Nakai, S},
  journal={Optics communications},
  volume={91},
  number={1-2},
  pages={9--12},
  year={1992},
  publisher={Elsevier}
}

@article{giacone1995angular,
  title={Angular dependence of stimulated Brillouin scattering in homogeneous plasma},
  author={Giacone, RE and McKinstrie, CJ and Betti, R},
  journal={Physics of Plasmas},
  volume={2},
  number={12},
  pages={4596--4605},
  year={1995},
  publisher={American Institute of Physics}
}

@incollection{goodman1975statistical,
  title={Statistical properties of laser speckle patterns},
  author={Goodman, Joseph W},
  booktitle={Laser speckle and related phenomena},
  pages={9--75},
  year={1975},
  publisher={Springer}
}

@article{taylor2006gaussian,
  title   = {A Gaussian Kinematic Formula},
  author  = {Taylor, Jonathan E.},
  journal = {The Annals of Probability},
  volume  = {34},
  number  = {1},
  pages   = {122--158},
  year    = {2006},
  doi     = {10.1214/009117905000000594}
}

\end{document}